\documentclass[nofootinbib,showpacs,preprint,preprintnumbers,amsmath,amssymb]{revtex4}
\usepackage{graphicx}
\usepackage{epsfig}
\usepackage{dcolumn}
\usepackage{bm}
\usepackage{amsmath}
\usepackage{color}
\usepackage{mathrsfs}
\usepackage{multirow}

\newcommand{\gsim}{\gtrsim}
\newcommand{\lsim}{\lesssim}

\newcommand{\no}{\nonumber}
\newcommand{\mc}{\mathcal}

\newcommand{\m}{\mathcal{M}}
\newcommand{\br}{{\rm Br}}
\newcommand{\hc}{{\rm H.c.}}

\newcommand{\gev}{{\;{\rm GeV}}}
\newcommand{\tev}{{\;{\rm TeV}}}
\newcommand{\beq}{\begin{equation}}
\newcommand{\eeq}{\end{equation}}
\newcommand{\bea}{\begin{eqnarray}}
\newcommand{\eea}{\end{eqnarray}}
\newcommand{\barr}{\begin{array}}
\newcommand{\earr}{\end{array}}
\newcommand{\bc}{\begin{center}}
\newcommand{\ec}{\end{center}}
\newcommand{\bit}{\begin{itemize}}
\newcommand{\eit}{\end{itemize}}
\newcommand{\ben}{\begin{enumerate}}
\newcommand{\een}{\end{enumerate}}

\newcommand{\sg}{\sigma}

\newcommand{\gm}{\gamma}

\newcommand{\lm}{\lambda}

\newcommand{\hobs}{h_{125}}

\newcommand{\mch}{m_{H^\pm}}
\newcommand{\tb}{t_\beta}
\newcommand{\cb}{c_\beta}
\renewcommand{\sb}{s_\beta}

\newcommand{\cba}{c_{\beta-\alpha}}
\newcommand{\sba}{s_{\beta-\alpha}}
\newcommand{\yh}{\hat{y}}
\newcommand{\gh}{\hat{g}}

\newcommand{\tR}{\widetilde{R}}

\newcommand{\rr}      {{\gamma\gamma}}

\newcommand{\ttau}      {{\tau^+\tau^-}}
\newcommand{\ttop}      {{t\bar{t}}}
\newcommand{\bb}      {{b \bar{b}}}

\newcommand{\hsm}      {{h_{\rm SM}}}

\renewcommand\epsilon{\varepsilon}

\begin{document}
\preprint{CTPU-15-07}
\title{Higgs potential and hidden light Higgs scenario\\ in two Higgs doublet models
}
\author{Sanghyeon Chang$^a$, Sin Kyu Kang$^b$, Jong-Phil Lee$^c$,
and Jeonghyeon Song$^d$}

\affiliation{
$^a$Center for Theoretical Physics of the Universe, IBS, Daejeon 305-811, Korea\\
$^b$School of Liberal Arts, Seoul-Tech, Seoul 139-743, Korea\\
$^c$School of Electrical Engineering and 
BK21 Plus Humanware Information Technology,
Korea University, Seoul 136-713, Korea\\
$^d$School of Physics,
Konkuk University, Seoul 143-701, Korea
}

\begin{abstract}
In two Higgs doublet models, there exists 
an interesting possibility, the hidden light Higgs scenario, 
that the discovered SM-like Higgs boson is the heavier \textit{CP}-even
Higgs boson $H^0$ and the lighter \textit{CP}-even
$h^0$ has not been observed yet in any experiment.
We study the current status of this scenario in Types I, II, X, and Y,
through the scans of the parameters
with all relevant theoretical and experimental constraints.
We employ not only the most up-to-date Higgs signal strength measurements
with the feed-down effects,
but also all the available LHC exclusion limits from heavy Higgs searches.
Adjusting the heavier $H^0$ to the 125 GeV state 
while hiding the lighter $h^0$ from the LEP Higgs search
prohibits the extreme decoupling limit:
there exist upper bounds on the masses of the pseudoscalar $A^0$ and
the charged Higgs $H^\pm$ below about $600\gev$.
In addition, the $Z_2$ symmetry
is shown to be a good approximate symmetry
since the soft $Z_2$ symmetry breaking parameter $m_{12}^2$ should be less than 
about $(45\gev)^2$.
Most interestingly,
a few parameters in the Higgs potential
and the related Higgs triple and quartic couplings
are shown to be meaningfully constrained by the current data.
The double Higgs-strahlung process at an $e^+ e^-$ collider is also studied.
\end{abstract}


\maketitle


\section{Introduction}
The observation of a Higgs boson with mass around 125 GeV at the LHC~\cite{Higgs:discovery:2012}
completes the journey in the standard model (SM):
the electroweak symmetry breaking mechanism is uncovered;
the mass generation of subatomic particles 
is most economically explained;
the Higgs boson mass, the last unknown parameter in the SM,
is precisely measured~\cite{Higgs mass}.
The current LHC Higgs data imply that the observed 125 GeV state $\hobs$ is
very similar to the SM Higgs 
boson~\cite{Aad:2014eha,Aad:2014lma,Khachatryan:2014ira,2HDM:ours,2HDM:ours2}.
Nevertheless there are some vital clues that this is not the end of the road.
We do not expect that the ultimate theory of particle physics is the SM which
suffers from the gauge hierarchy problem and has no solution to account for
95$\%$ of the energy of the Universe.
New physics is inevitable.

Apart from Occam's Razor, there is no reason 
for prohibiting additional Higgs doublets.
Many new physics models contain at least two Higgs doublets,
and thus additional Higgs bosons.
This extension of the Higgs sector is a good direction toward new physics beyond the SM.
The LHC Higgs data on the $125\gev$ state
may play the role of a compass to show the direction.
In addition, both ATLAS and CMS collaborations provide significant exclusion limits
from the null results in the searches for the heavy neutral and charged Higgs bosons.
In the ordinary setup where  the observed is the lightest \textit{CP}-even neutral Higgs boson $h^0$,
the compass naturally points to the decoupling limit~\cite{decoupling}
where the other Higgs states are very heavy.
The phenomenology of the decoupling limit generically mimics 
that of the SM.
Even if the current experimental status, 
the SM-like 125 GeV state without any signal of other Higgs bosons,
might keep in the future,
the verification or invalidation of a specific new physics model 
will be postponed till the next generation collider.

If the observed 125 GeV state is  a \emph{heavier} \textit{CP}-even neutral Higgs boson $H^0$,
however,
the LHC Higgs data 
play a much more significant role in characterizing a specific model.
Adjusting the heavier $H^0$ to $\hobs$
as well as hiding the lighter $h^0$ from low energy experiment data 
constrain the new physics model strongly.
We call this possibility the \emph{hidden light Higgs scenario}.
If the current LHC data can specify the Higgs potential
in this scenario thanks to the expected strong constraints,
it will give important implications 
on the dynamics of the electroweak phase transition~\cite{Noble:2007kk,Katz:2014bha},
and the measurement of the cubic and quartic self-couplings of Higgs bosons 
in the future collider~\cite{Battaglia:2001nn,Beyer:2006hx}.

As the simplest extension of the SM Higgs sector,
we consider a two Higgs doublet model (2HDM)~\cite{2hdm:review}
with \textit{CP} invariance and softly broken $Z_2$ symmetry~\cite{Z2sym}.
There exist five physical Higgs bosons,
the light \textit{CP}-even scalar $h^0$,
the heavy \textit{CP}-even scalar $H^0$, the \textit{CP}-odd pseudoscalar $A^0$,
and two charged Higgs bosons $H^\pm$.
The general Higgs potential has 7 parameters.
According to the $Z_2$ charges of the SM quarks and leptons,
there exist
four types of 2HDM: Type I, Type II, Type X, and Type Y~\cite{Aoki,Akeroyd:1996he}.
In the normal setup of $h^0=\hobs$,
there are extensive studies on the global fit analysis
of the Higgs signal strengths as well as the phenomenology of the other heavy Higgs bosons~\cite{Chen:2013kt,Chiang:2013ixa,Grinstein:2013npa,Eberhardt:2013uba,Belanger:2013xza,Cheung:2013rva,Celis:2013ixa,Wang:2013sha,Kanemura:2014bqa,Ferreira:2014sld,Broggio:2014mna,Chowdhury:2015yja}. 
The hidden light Higgs scenario is also naturally accommodated 
in the 2HDM~\cite{Wang:2014lta,2HDM:ours,Kanemura:2014dea,Bernon:2014nxa,Coleppa:2013dya,Cervero:2012cx,deVisscher:2009zb,Ferreira:2014dya}:
$H^0$ is the $125\gev$ state and $h^0$ has not been observed yet. 

\begin{table}
\caption{\label{table:summary:constraint}Summary of constraints.}
  {\renewcommand{\arraystretch}{1.1} 
  \begin{tabular}{|@{~}l@{~}|@{~}l@{~}|}
  \hline
Theoretical stability & $\cdot$ $V_H$ to be bounded below\\
\multirow{1}{*}{(yellow) }	& $\cdot$ Unitarity\\
	& $\cdot$ Perturbativity of quartic couplings\\ \hline
Pre-LHC bounds & $\cdot$ LEP bounds on $h^0$ and $H^\pm$ \\
 \multirow{1}{*}{(green)}   & $\cdot$ $\Delta\rho$ in the electroweak precision data\\
	& $\cdot$ FCNC like $\Delta M_{B_d}$ and  $b\to s \gm$ \\ 
	& $\cdot$ Top quark decay into $H^\pm$\\ \hline
LHC bounds & $\cdot$ $m_H = 125\gev$\\ 
\multirow{1}{*}{(red)}    & $\cdot$ LHC search for $H^\pm$ via $pp\to \ttop H^\pm$ followed by $H^\pm \to \tau\nu$\\
	& $\cdot$ LHC search for $A^0$ via $gg \to A^0 \to \rr,\ttau$ and $\bb\to A^0 \to\ttau$\\
	& $\cdot$ Global $\chi^2$ fit to the LHC Higgs data including \\
    &  \qquad (i) Additional decay channels of $H^0 \to h^0 h^0, A^0 A^0, H^+ H^-, W^\pm H^\mp$\\
    & \qquad (ii) the ``feed-down"\\	
	 \hline
\end{tabular}
}
\end{table}

A comprehensive study of the current status of the 2HDM
Type I and Type II by including the heavy Higgs search data
was first performed in Ref.~\cite{Dumont:2014wha}.
Similar comprehensive studies in other setup 
such as very light Higgs bosons~\cite{Bernon:2014nxa}
or the minimal supersymmetric standard model~\cite{Bhattacherjee:2015sga}
were followed up.
We extend the study, focusing on the 
question of how much the current data constrain the Higgs \emph{potential}
in the hidden light Higgs scenario.
To answer the question,
we consider more extended constraints than in Ref.~\cite{Dumont:2014wha},
particularly those from the LHC heavy Higgs searches.
We classify the theoretical and phenomenological constraints
into three categories:
theoretical bounds, pre-LHC bounds,
and LHC bounds.

The details of each step are summarized in Table \ref{table:summary:constraint}.
The theoretical bounds demand the boundedness of the Higgs potential~\cite{Ivanov:2006yq},
unitarity~\cite{Arhrib:2000is,Branco:2011iw}, and perturbativity.
The ``pre-LHC" bounds include
the LEP bounds on $h^0$~\cite{LEP,LEP2} and $H^\pm$~\cite{PDG:2014}, 
$\Delta\rho$ in the electroweak precision data~\cite{Higgs:Hunters:Guide,Chankowski:1999ta},
the flavor changing neutral current (FCNC) data such as
$\Delta M_{B_d}$ and $b\to s \gm$~\cite{bsr:2hdm:LO,Stal},
and the top quark decay of $t \to H^+ b$~\cite{Tevatron:cH:top:decay}.
The ``LHC" bounds are based not only on the Higgs signal strength measurements
from LHC8,
but also on the exclusion limits from
all the heavy Higgs searches
including $H^\pm \to \tau^\pm \nu_\tau$~\cite{ATLAS:t2bcH:cH2taunu,CMS:t2bcH:cH2taunu},
$H^+ \to c\bar{s}$~\cite{CMS:cH2cs},
$gg \to A^0 \to \rr$~\cite{Aad:2014ioa,CMS:2014onr},
$gg \to A^0 \to  \ttau$~\cite{Aad:2014vgg,Khachatryan:2014wca},
and
$\bb \to A^0 \to  \ttau$~\cite{Aad:2014vgg,Khachatryan:2014wca}.
For the $125\gev$ state data,
we perform the comprehensive global $\chi^2$ analysis
by including $H^0 \to h^0 h^0$ and $H^0 \to A^0 A^0$
as well as ``feed-down" (FD) contributions
from the production of heavier Higgs bosons 
through their decay into $H^0$~\cite{Dumont:2014wha,Arhrib:2013oia},
not by just limiting the FD signal strength value.

We shall show that in all four types,
the hidden light Higgs scenario
is consistent with the data, as good as the SM.
In addition, the survived parameter points
have many interesting implications.
Major ones are as follows:
(i) there exist \emph{upper} bounds on the heavy Higgs bosons like
$m_{A,H^\pm} \lsim 600\gev$;
(ii) the soft $Z_2$ symmetry breaking term $m_{12}^2$
is strongly constrained to be below about $(45\gev)^2$ in most cases;
(iii) in Type I and X, a light $m_A$ (as low as 20 GeV) and $\mch$ 
(as low as  100 GeV)
are allowed;
(iv) the triple Higgs coupling $g_{HHH}$ 
is very like the SM Higgs triple coupling, and $g_{hHH}$ is 
similar to the SM value;
(v) the quartic Higgs couplings $g_{HHHH}$ and $g_{hHHH}$
are similar to the SM value.
Inspired by the almost fixed $g_{HHH}$ and $g_{hHH}$,
we study the double Higgs􏲄-strahlung at an $e^+ e^-$ collider,
$e^+ e^- \to Z^0 H^0 H^0$,
which can be highly enhanced in some parameter space 
where the resonance production of $A^0$ is allowed. 
These are our main results.

The paper is organized as follows.
In Sec.~\ref{sec:review},
we briefly review the 2HDM.
Section \ref{sec:constraints}
summarizes the constraints.
Section \ref{sec:results} presents our results,
the allowed parameter space stage by stage.
In Sec.~\ref{sec:ee2ZHH},
we show the Higgs triple and quartic couplings in the allowed parameter space,
and study
the future prospect of $e^+ e^-\to Z^0 H^0 H^0$. 
Section \ref{sec:conclusions} contains our conclusions.

\section{Brief Review of the 2HDM}
\label{sec:review}
A 2HDM~\cite{2hdm:review} introduces
two complex $SU(2)_L$ Higgs doublet scalar fields, $\Phi_1$ and $\Phi_2$.
Both $\Phi_1$ and $\Phi_2$ develop nonzero vacuum expectation values  as
$
\langle \Phi_{1,2} \rangle = \left (0, v_{1,2}/\sqrt{2} \right)^T$,
which generate the electroweak symmetry breaking.
When parametrising $\tb =v_2/v_1$,
one linear combination $H_1 = \cb \Phi_1 + \sb \Phi_2$
has nonzero vacuum expectation value of $v =\sqrt{v_1^2+v_2^2}=246\gev $,
while its orthogonal combination $H_2 = -\sb \Phi_1 +\cb \Phi_2$
acquires zero vacuum expectation value.
For simplicity of notation,  we take $s_x=\sin x$, $c_x = \cos x$, and $t_x = \tan x$.
We define the fluctuation fields about the minima $v_1$ and $v_2$ as
\bea
\Phi_i = \left( \begin{array}{c} \phi_i^+ \\
\dfrac{v_i +  \rho_i + i \eta_i }{ \sqrt{2}} 
\end{array} \right),
\quad i=1,2
\,.
\eea
In order to avoid FCNC at the lowest order,
a discrete $Z_2$ symmetry is imposed, under which $\Phi_1 \to \Phi_1$
and $\Phi_2 \to -\Phi_2$~\cite{Z2sym}.

The most general potential with \textit{CP} invariance and softly broken $Z_2$ symmetry is  
\bea
\label{eq:V}
V= && m^2 _{11} \Phi^\dagger _1 \Phi_1 + m^2 _{22} \Phi^\dagger _2 \Phi_2 
-m^2 _{12} ( \Phi^\dagger _1 \Phi_2 + \hc) \no \\
&& + \frac{1}{2}\lambda_1 (\Phi^\dagger _1 \Phi_1)^2 
+ \frac{1}{2}\lambda_2 (\Phi^\dagger _2 \Phi_2 )^2 
+ \lambda_3 (\Phi^\dagger _1 \Phi_1) (\Phi^\dagger _2 \Phi_2) 
+ \lambda_4 (\Phi^\dagger_1 \Phi_2 ) (\Phi^\dagger _2 \Phi_1) \no \\
&& + \frac{1}{2} \lambda_5 
\left[
(\Phi^\dagger _1 \Phi_2 )^2 +  \hc
\right] . \label{eq:potential}
\eea
The \textit{CP} invariance requires all of the parameters to be real,
and $m_{12}^2$ breaks the $Z_2$ symmetry softly.
Note that the $m_{12}^2$ parameter can be negative.
Using the tadpole condition, $m_{11}$ and $m_{22}$
can be written in terms of $v$, $\tb$, and 
$\lm_{1,\cdots,5}$.
The Higgs potential has 7 free parameters of 
$m_{12}^2$, $\tb$, and $\lm_{1,\cdots,5}$.

The charged Higgs boson $H^\pm$ is a linear combination of $\phi^\pm_1$ and $\phi^\pm_2$,
and the pseudoscalar $A^0$ is a linear combination of $\eta_1$ and $\eta_2$.
Their orthogonal states are Goldstone modes $G^\pm$ and $G^0$, respectively.
And their masses are
\bea
\label{eq:mA:mch}
\mch^2 =  \frac{m^2_{12} }{\cb\sb} 
- (\lambda_4 + \lambda_5 ) v^2,\quad
m^2_A = \frac{m^2_{12} }{\cb\sb} 
-2 \lm_5 v^2.
\eea
The physical \textit{CP}-even Higgs bosons $h^0$ and $H^0$  
are
obtained through the diagonalization of
the mass squared matrix $\m^2_{0}$ with the mixing angle $\alpha$, given by
\bea
\m^2_{0} = 
\left ( \barr{cc} \m^2 _{11} & \m^2 _{12} \\ \m^2 _{12} & \m^2 _{22} \earr \right )
\eea
where
\bea
\label{eq:Msqij}
\mc{M}^2 _{11} &=&m^2_{12} \tb^2 + \frac{\lm_1 v^2}{1+\tb^2},
\quad \mc{M}^2 _{22} =\frac{m^2_{12} }{ \tb^2} + \lm_2 v^2 \frac{\tb^2}{1+\tb^2}
\\ \no
\mc{M}^2 _{12} &=& -m^2_{12} + \lm_{345} v^2 \frac{\tb}{1+\tb^2},
\eea
where $\lm_{345}=\lambda_3+\lambda_4+\lambda_5$.
The masses of neutral Higgs $m_h$ and $m_H$ are
\bea
\label{eq:mhHmass}
m^2 _{H,h} &=& \frac{1}{2}\Big  [ \mc{M}^2 _{11} + \mc{M}^2 _{22} \pm \sqrt{(\mc{M}^2_{11} - \mc{M}^2 _{22} )^2 + 4 (\mc{M}^2 _{12})^2 }~ \Big ]. 
\eea

The SM Higgs field, which corresponds to $H_1$,
becomes
\bea
h_{\rm SM} = \sba h^0 + \cba H^0.
\eea
If $\cba=1$ and $m_H = 125\gev$,
$H^0$ has the same properties as the SM Higgs boson.
This is called the alignment limit~\cite{alignment}: 
\bea
\label{eq:alignment}
\hbox{The alignment limit for $H^0 = h_{\rm SM}$: }&& \cba=1.
\eea
As shall be shown, the allowed parameters by all the constraints
are distributed around the alignment limit.
The alignment limit maximizes or minimizes some
triple couplings of Higgs bosons
with weak gauge bosons or other Higgs bosons.
We classify them into 
two categories, one proportional to $\sba$ and the other proportional to $\cba$:
\bea
\label{eq:sba:cba}
 \sba &:& g_{hW^+ W^-}, \quad g_{hZZ}, \quad g_{ZAH}, \quad g_{W^\pm H^\mp H},
\\ \no
\cba &:& g_{HW^+ W^-}, \quad g_{HZZ}, \quad g_{ZAh}, \quad g_{W^\pm H^\mp h},
\quad g_{Hhh}.
\eea
In the hidden light Higgs scenario,
the couplings proportional to $\sba$ vanish in the alignment limit.

Yukawa couplings of Higgs bosons are different according to the 2HDM type.
Focusing on the 125 GeV state $H^0$,
we present the normalized Yukawa couplings by the SM values, $\yh^{H}_{uu,dd,\ell\ell}$,
in terms of $\cba$ and $\sba$:
\bea
\label{eq:Yukawa:couplings}
\begin{array}{l|cc}
                & ~~~~~\cba - \dfrac{\sba}{\tb}~~~~~ & ~~~~~\cba+\tb\sba~~~~~\\
                \hline
\hbox{Type I } & \yh^H_{uu} , \yh^H_{dd},\yh^H_{\ell\ell} & \\
\hbox{Type II } & \yh^H_{uu} & \yh^H_{dd},\yh^H_{\ell\ell} \\
\hbox{Type X } & \yh^H_{uu} , \yh^H_{dd}   &\yh^H_{\ell\ell}  \\
\hbox{Type Y } & \yh^H_{uu},\yh^H_{\ell\ell} & \yh^H_{dd} \\
\end{array}
\eea
Note that in the exact alignment limit ($\cba=1$), all of the Yukawa couplings 
for $H^0$ are the same as in the SM.
There are two kinds of deviation from the alignment limit:
one is proportional to $\tb$, and the other to $1/\tb$.
Since FCNC constrains $\tb\gsim 1$,
those proportional to $\tb\sba$ yield 
much larger deviation from the SM Yukawa coupling.
As shown in Eq.~(\ref{eq:Yukawa:couplings}),
Type I has common Yukawa couplings,
which have the $\tb$-suppressed deviation from the alignment.
On the while, Type II has the $\tb$-enhanced deviation 
for the down-type quark
and charged lepton Yukawa couplings.
As shall be shown later, Type II is most strongly constrained by the current LHC Higgs data.

\section{Constraints}
\label{sec:constraints}

We constrain the hidden light Higgs scenario in the 2HDM  by sequentially taking
three steps.
The first step (yellow) is to apply  theoretical conditions,
the second (green) is to use all reliable experimental constraints before the LHC data,
and the last step (red) is to include the LHC Higgs data that consist of the
observation of the $125\gev$ state
as well as the exclusion limits from the searches 
for the other Higgs bosons heavier than 125 GeV.
In what follows, each coloured point (yellow, green, or red) represents
the surviving parameters at 95\% C.L.,  up to the corresponding step.
For example,
the green points satisfy the theoretical and pre-LHC bounds.
We summarise the constraints in Table \ref{table:summary:constraint}.

\subsection{Theoretical constraints}
\label{subsec:theoretical}
\textit{(i) The Higgs potential to be bounded from below}:
As proven in Ref.~\cite{Ivanov:2006yq},
the scalar potential in Eq.~(\ref{eq:V}) is bounded from below 
if and only if the following
conditions are satisfied:
\bea
\lambda_1 > 0, \quad 
\lambda_2 > 0, \quad \lambda_3 > -\sqrt{\lambda_1 \lambda_2}, \quad  
\lambda_3 + \lambda_4 - |\lambda_5| > - \sqrt{\lambda_1 \lambda_2}.  \label{eq:stability} 
\eea

\textit{(ii) Unitarity}:
Tree level perturbative unitarity requires for the absolute values of the followings 
to be less than $8\pi$~\cite{Arhrib:2000is,Branco:2011iw}:
\bea
a_\pm & = & \frac{3}{2} (\lambda_1+\lambda_2)\pm \sqrt{\frac{9}{4}(\lambda_1-\lambda_2)^2+(2\lambda_3+\lambda_4)^2},\\ \no
b_\pm & = &\frac{1}{2}\left ( \lambda_1+\lambda_2 \pm \sqrt{(\lambda_1-\lambda_2)^2+4\lambda_4^2} \right ),\\ \no
c_\pm & = & \frac{1}{2}\left ( \lambda_1+\lambda_2 \pm \sqrt{(\lambda_1-\lambda_2)^2+4\lambda_5^2} \right ),\\ \no
f_+ &=&  \lambda_3 +2\lambda_4+3\lambda_5, \quad f_- = \lambda_3 +\lambda_5, \quad f_{1}= \lambda_3 +\lambda_4,\\ \no
e_{1}&=& \lambda_3+2\lambda_4-3\lambda_5, \quad e_{2}=\lambda_3-\lambda_5, \quad p_1 = \lambda_3-\lambda_4.
\eea

\textit{(iii) Perturbativity}:
We first demand the bare quartic couplings in the Higgs potential to satisfy 
the perturbativity as
\bea
\label{eq:perturbativity}
\left| \lm_{i} \right| < 4 \pi, \quad i=1,\cdots,5.
\eea
In addition, the magnitudes of the quartic couplings among physical Higgs states like 
$g_{\varphi_i \varphi_j\varphi_k \varphi_\ell} (\phi_i=h^0, H^0, A^0, H^\pm)$
are required to be smaller than $4\pi$.

Here we do not require that the 2HDM vacuum should be 
the global minimum of the potential~\cite{Barroso:2013awa}, because
the existence of a false vacuum (local minimum) is acceptable if its lifetime 
is longer than
the age of the Universe.  
Unquestionably if the lifetime of a false vacuum is shorter,
the corresponding parameter space 
should be excluded. 
Since the calculation of the 2HDM vacuum lifetime is beyond the scope of this study,
we take a conservative stance to ignore the global minimum condition.

\subsection{pre-LHC bounds}
\label{subset:pre-LHC bounds}

\textit{(i)  The LEP  bounds on $h^0$ and $H^\pm$}:
One of the most direct channels to probe a light Higgs boson 
with mass below 120 GeV
is the Higgs-strahlung at the LEP.
We use the strongest upper bound 
on the event rate of $e^+ e^- \to Z^0 h^0 \to Z^0 jj$~\cite{LEP,LEP2}.
Another important result from the LEP
is the direct production limit on the charged Higgs boson mass as~\cite{PDG:2014}:
\bea
m_{H^\pm} \geq 80\gev.
\eea

(ii) \textit{$\Delta\rho$ in the electroweak precision data}:
The $\Delta\rho$ parameter from the electroweak precision measurement
has additional contributions in the 2HDM through the heavy neutral Higgs bosons
($A^0$ in the hidden light Higgs scenario)
as well as
the charged Higgs bosons~\cite{Higgs:Hunters:Guide,Chankowski:1999ta}.
With the observed Higgs boson mass, 
its global fit result has been improved significantly as~\cite{PDG:2014}
\begin{equation}
\Delta\rho= 0.00040\pm 0.00024\,.
\end{equation}
It is known that the new contribution is 
suppressed if $m_A \simeq \mch$ or $m_H \simeq \mch$~\cite{Gerard:2007kn,Song:2014lua}.
As shall be discussed,
the  $m_H \simeq \mch \simeq 125\gev$ case is prohibited by the FCNC constraints
in Types II and Y. 

\textit{(iii) FCNC processes such as $\Delta M_{B_d}$ and  $b\to s \gm$}:
In the 2HDM, the charged Higgs boson contributes to various FCNC processes 
through the loop.
We consider two sensitive FCNC processes, $b \to s \gm$~\cite{bsr:2hdm:LO,Stal} and 
$\Delta M_{B_d}$~\cite{Heavy Flavor Averaging Group,Stal}. 
$\Delta M_{B_d}$  excludes small $\tb$ region for all types,
while $b \to s \gm$ further excludes 
the light charged Higgs mass region in Type II and Type Y.
Other processes such as $\epsilon_K$~\cite{epsilonK:Jung} and $R_b$~\cite{Rb:exp,Rb:2hdm}
impose weaker constraints~\cite{EWfit}.
We do not consider the measurements  of  
$R(D^{(\ast)})\equiv \mbox{Br}(\bar{B} \rightarrow D^{(\ast)} \tau^{-} \bar{\nu}_{\tau})/
\mbox{Br}(\bar{B} \rightarrow D^{(\ast)} l^{-} \bar{\nu}_{l})$ from BaBar~\cite{Lees:2012xj}\footnote{ Recently, LHCb~\cite{Aaij:2015yra} reported $2.1\sigma$ excess of $R(D^{\ast})$ over the SM predictions, and
Belle~\cite{Huschle:2015rga} presented a new measurements 
of $R(D)$ and $R(D^*)$ which are not significantly deviated 
from both the SM prediction and the measured values at BaBar~\cite{Lees:2012xj} and LHCb.}.


\textit{(iv) Bound from $t \to b H^+$}:
A light charged Higgs boson could have appeared in the top decay into $b H^+$
if kinematically allowed.
We include the 
Tevatron search results of the upper bounds on 
$\br(t \to b H^+)$~\cite{Tevatron:cH:top:decay}.

\subsection{LHC bounds}
\textit{(i) Higgs mass bounds}:
Both ATLAS and CMS collaborations measured the Higgs boson mass 
with high precision~\cite{Aad:2014aba,Khachatryan:2014jba}.
The combined result is~\cite{Aad:2015zhl}
\bea
\label{eq:Higgs:mass}
m_H = 125.09 \pm 0.21 (\hbox{stat}) \pm 0.11 (\hbox{syst}) \gev .
\eea
We demand that $m_H$ be within $2\sigma$.
In addition, we exclude the degenerate cases of $m_H = m_A$ and $m_h = m_H$ 
in order to avoid the possible contributions from $h^0$ or $A^0$ to the observed Higgs signal strengths.
Both $m_h$ and $m_A$ should lie outside $m_H$ at $2\sigma$.

\textit{(ii) LHC search for the charged Higgs boson}:
The search strategy of the charged Higgs boson
at the LHC is different according to its mass.
For $\mch>m_t+m_b$,
the main decay mode is into $t \bar{b}$~\cite{CMS:cH2tb}.
For lighter charged Higgs boson than the top quark,
two decay channels are searched, 
$H^\pm \to \tau^\pm \nu_\tau$~\cite{ATLAS:t2bcH:cH2taunu,CMS:t2bcH:cH2taunu}
and $H^+ \to c\bar{s}$~\cite{CMS:cH2cs}.
Since the direct production of the charged Higgs boson is very small,
the bound for the charged Higgs boson is weak in general.
The strongest bound is from $pp \to t\bar{t} \to \bb H^\pm W^\mp$,
followed by $H^\pm \to \tau^\pm\nu$.
We include the upper bounds on $\br(t \to H^+ b) \times \br(H^+ \to \tau^+ \nu_\tau)$.

\textit{(iii) LHC search for $A^0$}:
In the hidden light Higgs scenario,
only $A^0$ is the heavy neutral Higgs boson.
Up to now, there are no significant excesses in the heavy neutral Higgs search,
which provides the exclusion limit.
We include the $gg \to A^0 \to \rr$~\cite{Aad:2014ioa,CMS:2014onr},
$gg \to A^0 \to  \ttau$~\cite{Aad:2014vgg,Khachatryan:2014wca}, and
$\bb \to A^0 \to  \ttau$~\cite{Aad:2014vgg,Khachatryan:2014wca}.
Note that the $W^+W^-$ and $Z^0Z^0$ decay channels
are not relevant for the pseudoscalar $A^0$.
Another important decay channel is into $\ttop$,
which is dominant if $m_A > 2 m_t$ and $\tb \lsim 10$~\cite{Song:2014lua}.
Although both ATLAS and CMS collaborations 
reported the $\ttop$ resonance search results~\cite{heavyH:tt},
the interpretation  is very challenging at a hadron collider
because the interference with the QCD continuum background
causes various shapes in the $\ttop$ invariant mass 
distribution~\cite{Jung:2015gta,Dicus:1994bm,Djouadi:2015jea}.
Since the interference effects have not been included yet in the experiment analysis,
we do not consider this $\ttop$ channel.

\begin{table}
\centering
\caption{\label{tab:R}Summary of the LHC Higgs signal strengths at 7 and 8 TeV.}
\begin{tabular}{|c|l|l|}
\hline
Production &  \multicolumn{1}{c|}{ATLAS}  & \multicolumn{1}{c|}{CMS}\\
\hline
${gg {\rm F}+t\bar{t}h}$ & ~${\widetilde{R}\,}_{\gamma\gamma}^{gg {\rm F}} = 1.32\pm 0.38$ \cite{Aad:2014eha} ,
 ~${\widetilde{R}\,}_{\gamma\gamma}^{t\bar{t}h} = 1.3^{+2.6}_{-1.7}$ \cite{Aad:2014lma}
	      & ~${\widetilde{R}\,}_{\gamma\gamma}^{gg {\rm F}+t\bar{t}h} = 1.13^{+0.37}_{-0.31} $ \cite{Khachatryan:2014ira} \\
            & ~${\widetilde{R}\,}_{WW}^{gg \rm F} = 1.01^{+0.27}_{-0.20}$ \cite{ATLAS-CONF-2014-060} 
	      & ~${\widetilde{R}\,}_{WW}^{gg \rm F} = 0.74^{+0.22}_{-0.20}$ \cite{Chatrchyan:2013iaa}\\
            & ~${\widetilde{R}\,}_{ZZ}^{gg {\rm F}+t\bar{t}h} = 1.52^{+0.85}_{-0.65}$ \cite{ATLAS-CONF-2014-009}
	      & ~${\widetilde{R}\,}_{ZZ}^{gg {\rm F}+t\bar{t}h} = 0.80^{+0.46}_{-0.36}$ \cite{Chatrchyan:2013mxa} \\
            & ~${{\widetilde{R}\,}_{{\tau\tau}}^{gg \rm F} = 1.93^{+1.45}_{-1.15}}$ \cite{ATLAS-CONF-2014-061} 
	      & ~${{\widetilde{R}\,}_{{\tau\tau}}^{gg \rm F} = 0.93 \pm 0.42 }$ \cite{Chatrchyan:2014nva}\\
               &~${{\widetilde{R}\,}_{b \bar b}^{t\bar th} = 1.7 \pm 1.4}$ \cite{ATLAS-CONF-2014-011,Llacer:2014lna} & ~${\widetilde{R}\,}_{b \bar b}^{t\bar th} = 0.67^{+1.35}_{-1.33} $ \cite{Vernieri:2014wfa} \\
\hline
${{\rm VBF}+Vh}$   & ~${\widetilde{R}\,}_{\gamma\gamma}^{{\rm VBF}} = 0.8\pm0.7$ ,
	      & \\
 & ~${\widetilde{R}\,}_{\gamma\gamma}^{WH} = 1.0\pm 1.6$,   ${\widetilde{R}\,}_{\gamma\gamma}^{ZH} = 0.1^{+3.7}_{-0.1}$ \cite{Aad:2014eha}
	      & ~${\widetilde{R}\,}_{\gamma\gamma}^{{\rm VBF}+Vh} = 1.15 ^{+0.63}_{-0.58}$  \cite{Khachatryan:2014ira}\\
             & ~${\widetilde{R}\,}_{WW}^{\rm VBF} = 1.28^{+0.53}_{-0.45}$ \cite{ATLAS-CONF-2014-060} 
	      & ~${\widetilde{R}\,}_{WW}^{\rm VBF} = 0.60^{+0.57}_{-0.46}$, 
                     ${\widetilde{R}\,}_{WW}^{Vh} = 0.39^{+1.97}_{-1.87}$  \cite{Chatrchyan:2013iaa} \\
             & ~${\widetilde{R}\,}_{ZZ}^{{\rm VBF}+Vh} = 0.90^{+4.5}_{-2.0}$ \cite{ATLAS-CONF-2014-009}
	      & ~${\widetilde{R}\,}_{ZZ}^{{\rm VBF}+Vh} = 1.7^{+2.2}_{-2.1}$ \cite{Chatrchyan:2013mxa} \\
             & ~${{\widetilde{R}\,}_{{\tau\tau}}^{{\rm VBF}+Vh} = 1.24^{+0.58}_{-0.54}}$ \cite{ATLAS-CONF-2014-061} 
	      &~${{\widetilde{R}\,}_{{\tau\tau}}^{\rm VBF} = 0.94\pm0.41 }$,
                     ${{\widetilde{R}\,}_{{\tau\tau}}^{Vh} = -0.33\pm1.02}$ \cite{Chatrchyan:2014nva} \\
             & ~${{\widetilde{R}\,}_{b \bar b}^{{Vh}} = 0.51^{+0.40}_{-0.37}} $ \cite{Aad:2014xzb}
	      & ~${ {\widetilde{R}\,}_{b \bar b}^{\rm VBF} = 0.7\pm 1.4 }$
\cite{Vernieri:2014wfa}, ~${ {\widetilde{R}\,}_{b \bar b}^{Vh} = 1.0\pm 0.5 }$ \cite{Chatrchyan:2013zna}\

              \\ \hline
              \end{tabular}
\end{table}


\textit{(iv) the global fit to the LHC 125 GeV state data with the FD effects}:
The discovery of the Higgs boson is not based on a single observation of a 
resonance, 
but more than 200 channels.
Any new physics model should explain the whole LHC Higgs data,
which is commonly analysed through the global $\chi^2$ fit. 
We parameterize each signal rate by $R^{\tt production}_{\tt decay}$,
the ratio of the observed event rate to the SM expectation
in the specific channel,
and identify it with  the signal strength modifier $\mu\equiv \sg/\sg_{\rm SM}$.
In the 2HDM, $R$'s are not generally equal to one as in the SM.
The latest experimental values, denoted by $\tilde{R}$'s, 
are summarised in Table \ref{tab:R}.
We perform global $\chi^2$ fits of 7 model parameters 
to the observed Higgs signal strength $\tR_i$.

In the 2HDM,
there are three sources to deviate the signal strength value from one.
First, the effective couplings of $H^0$ with the SM particles
can be different from those in the SM,
which happens when $\cba \neq 1$.
Second, there are additional decay channels of $H^0$ into $h^0 h^0$,
$A^0 A^0$, $Z^0 A^0$, $W^\pm H^\mp$, and $H^+ H^-$.
Too large decay rates of new decay modes
enhance the total decay rate of $H^0$, which 
affects the observed signal strengths.
We include these decay rates to the global $\chi^2$ fit.
In the hidden light Higgs scenario,
the $H^0 \to h^0 h^0$ mode 
excludes considerable parameter space even in the alignment limit,
since the vertex is proportional to $\cba$.
On the contrary, the vertices of $Z^0$-$A^0$-$H^0$ and $W^\pm$-$H^\mp$-$H^0$
are proportional to $\sba$,
and thus $H^0 \to Z^0 A^0,W^\pm H^\mp$
are suppressed in the alignment limit: see Eq.~(\ref{eq:sba:cba}).
The LEP bounds of $m_{H^\pm} \geq 80\gev$ kinematically suppresses $H^0 \to H^+ H^-$.

The third source for the deviation is the FD effects~\cite{Dumont:2014wha,Arhrib:2013oia}:
the inclusive decay of heavy Higgs states into $H^0$ yields more events
in the $125\gev$ state, which renders the $H^0$ not to be SM-like even in the alignment 
limit. 
In the hidden light Higgs scenario,
dominant contribution to the FD effects is from the inclusive decay of $A^0$ into $H^0$.
The $H^\pm$ contribution is negligible since its production at the LHC is too small.
The general probability
of the inclusive production of $H^0$ from $A^0$ decay is~\cite{Dumont:2014wha}
\bea
\label{eq:prob:A:FD}
\mathcal{P}_{\rm FD} ( A^0 \to H^0 + X) &=& 
2 \br(A^0 \to H^+ H^-) \br(H^+ \to W^+ H^0)^2 + \br(A^0 \to Z^0 H^0 ) \\ \no
&&+ 2 \br(A^0 \to W^- H^+) \br(H^+ \to W^+  H^0)
\\ \no &&
+ 2 \br(A^0 \to H^+ H^-) \br (H^+ \to W^+ H^0) \left\{
1- \br (H^+ \to W^+ H^0)
\right\}.
\eea
As shall be shown in the next section,
the ``pre-LHC" constraints allow two kinds of regions 
in the $(\mch,m_A)$ parameter space.
One is 
$ \mch \simeq m_A$ in all four types, and the other allows
$\mch \approx 100\gev$.
Only the $\br(A^0 \to Z^0 H^0 )$ in Eq.~(\ref{eq:prob:A:FD}) is kinematically relevant.

We define new FD signal strengths as 
\bea
\label{eq:mu:FD}
\mu^{\rm FD:ZH}_{ii} = 
\frac{ \sg ( pp \to gg\to A^0) \br (A^0 \to Z^0 H^0)}{\sg(pp \to Z^0 \hsm)}
\times
\frac{\br(H^0 \to ii)}{\br(\hsm \to ii)},
\eea
where $i=\gm,W,Z,\tau,b$\footnote{Our definition has additional factor
of $\br(H^0 \to ii)/\br(\hsm \to ii)$ compared with the $\mu^{\rm FD}$
in Ref.~\cite{Dumont:2014wha}, which is the ratio of the FD production
to direct production in the 2HDM.}.
We add $\mu^{\rm FD:ZH}_{ii}$ to $R^{ZH}_{ii}$
and perform
the global $\chi^2$ fit to the observed $\tR$'s.

\section{Results}
\label{sec:results}
For the 7 parameters of $\lm_{1,\cdots,5}$, $\tb$, and $m_{12}^2$,
we randomly generate $2 \times 10^{10}$ points to scan over the following ranges:
\bea
&&\lm_{1,2} \in [0,4\pi],\quad \lm_{3,4,5} \in [-4\pi,4\pi],
\\ \no
&&\tb \in [1,50], \quad m_{12}^2 \in [-(2\tev)^2, (2\tev)^2].
\eea
We apply three steps of bounds:
\begin{description}
\item[Step-1 (yellow)] theoretical bounds;
\item[Step-2 (green)] the pre-LHC bounds;
\item[Step-3 (red)] the LHC bounds.
\end{description}
The detailed conditions are summarised in Table \ref{table:summary:constraint}.
In what follows, yellow, green, and red points denote the survived parameters
after Step-1, Step-2, and Step-3 bounds, respectively.
Note that the bound conditions are accumulatively applied.
The red points satisfy all the bounds.

\textit{(i) High reliability of the hidden light Higgs scenario}:
We find that the hidden light Higgs scenario is  
consistent with all the current data.
Out of $2\times 10^{10}$ parameter sets and at 95\% C.L.,
$2.2\times 10^{4}$ points survived in Type I,
$0.74\times 10^{4}$ in Type II, $1.1\times 10^4$ in Type X, and
$1.4\times 10^{4}$ in Type Y.
Limited but substantial parameter space is consistent with the current data.
Type I has the largest allowed parameter space while
Type II has the smallest.
The minimum values of $\chi^2$ per degree of freedom
are $0.40$, $0.51$, $0.51$, $0.50$ in Type I, II, X, and Y, respectively.
In the SM, it is $0.49$.
The best fit points in the 2HDM explain the current data at least as
good as the SM.

\begin{figure}[t!]
\centering
\includegraphics[width=7cm]{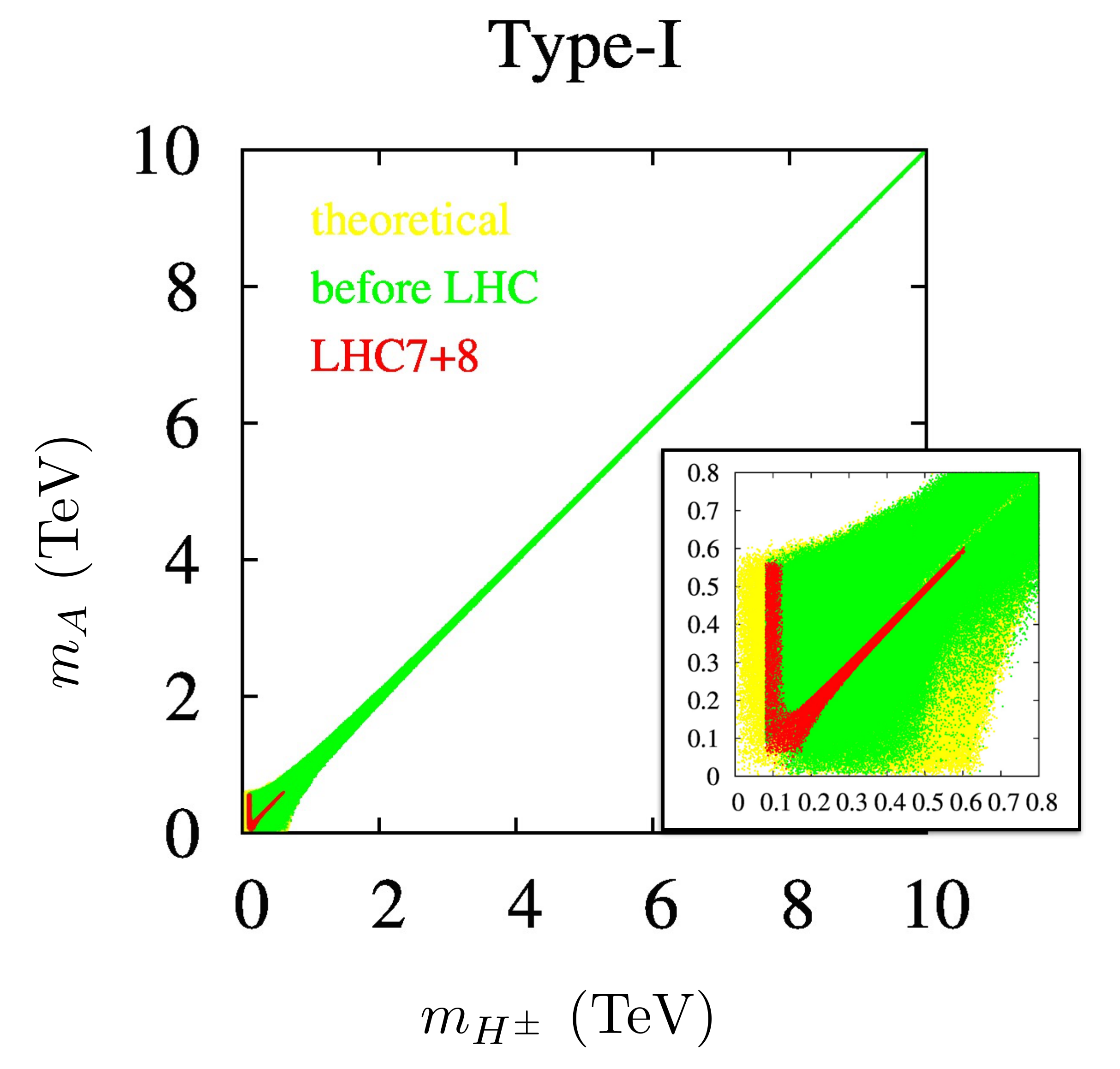}~~~~~
\includegraphics[width=7cm]{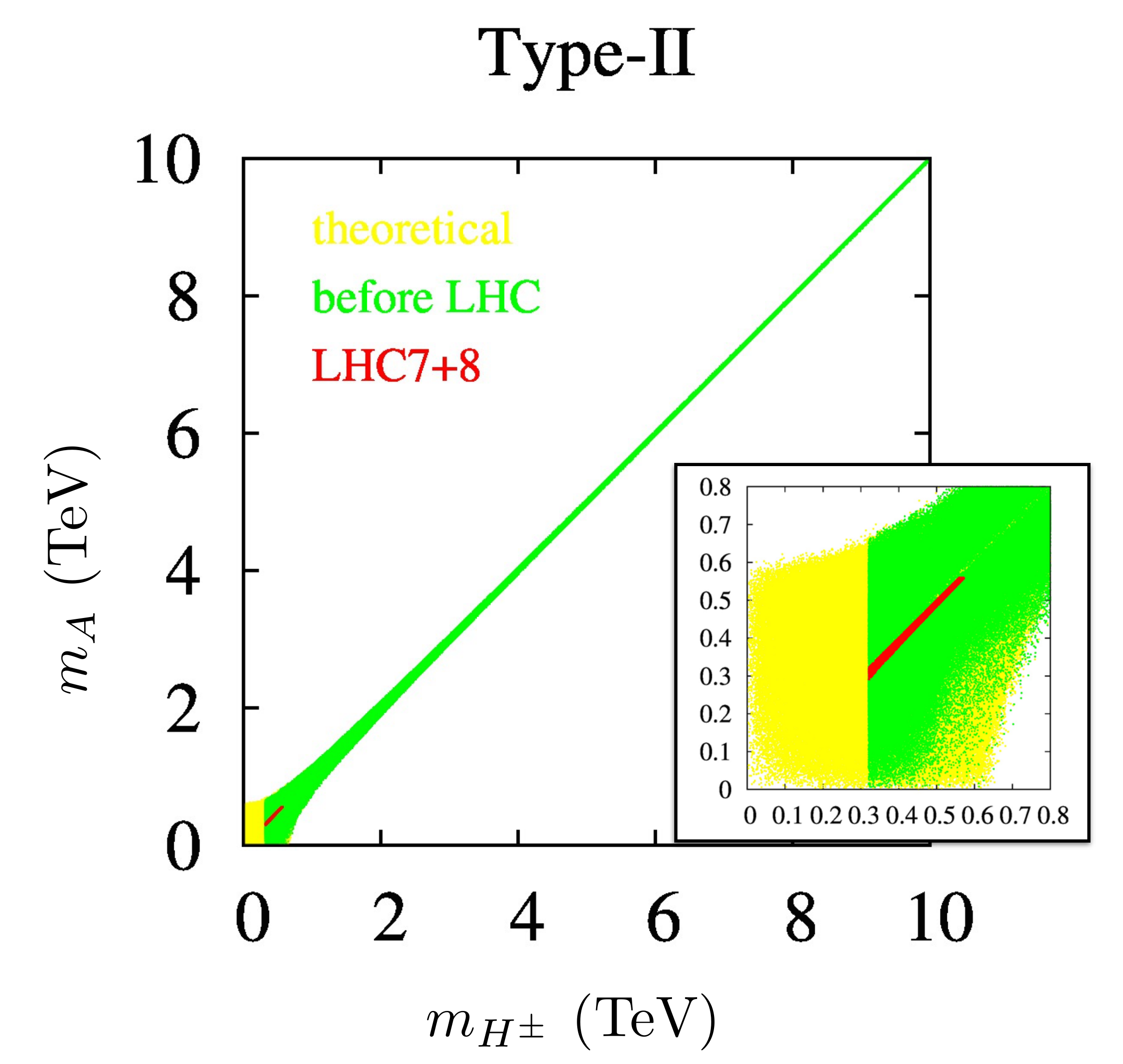}
\caption{\label{fig:mAmcH}
Constraints in the $m_A$ versus $\mch$ plane for Type I and Type II 2HDM.
Yellow points satisfy theoretical bounds.
Green points satisfy up to the pre-LHC constraints
while red points satisfy up to LHC constraints.
The details of the constraints are summarised in Table \ref{table:summary:constraint}.
 }
\end{figure}

\textit{(ii) Upper bounds on $m_A$ and $\mch$}:
A unique feature of the hidden light Higgs scenario is
the presence of the upper bounds on heavy Higgs boson masses, $m_A$ and $\mch$.
In the normal setup where $\hobs=h^0$,
the so-called decoupling limit can be always taken such that
all of the new particles are heavy enough to be beyond the LHC reach.
A safety zone for avoiding the invalidation of the model by the experimental data 
is guaranteed.
In the hidden light Higgs scenario, however,
matching the heavy $H^0$ to the 125 GeV state
constrains the other heavy Higgs bosons, $A^0$ and $H^\pm$.

Figure \ref{fig:mAmcH} presents 
the allowed $m_A$ and $\mch$ in Type I and Type II.
The results for Type X (Type Y) are very similar to Type I (Type II).
As summarised in Table \ref{table:summary:constraint},
yellow, green, and red points satisfy theoretical bounds, pre-LHC constraints, 
and LHC constraints, respectively.
Before the LHC data, 
a band-shaped region of $m_A \simeq \mch$ is allowed in the $(\mch, m_A)$ space.
For Type II, there exist a lower bound on $\mch$,
which is constrained by the flavor data, 
and a lower bound on $m_A$ due to the $\Delta\rho$.
There are no upper bounds on $m_A$ and $\mch$ at this stage.
Large masses up to $10\tev$ are possible,
which are dominantly from the $m_{12}^2$ terms in Eq.~(\ref{eq:mA:mch}).
The degeneracy of $m_A \approx \mch$
is also explained by the same $m_{12}^2$ terms of $m_A$ and $\mch$,
which suppresses the new contribution to the $\Delta\rho$.

When the LHC Higgs data are applied, 
the most of the green band region is excluded, leaving a very limited parameter space.
As shown in the magnified small mass region,
the LHC Higgs data put upper bounds on $m_A$ and $\mch$.
Both should be less than about $600\gev$.
The strongest bound is from the observed mass of the 125 GeV state
with high precision
in Eq.~(\ref{eq:Higgs:mass}).
We expect that at the LHC Run 2 both $A^0$ and $H^\pm$
are to be probed in most of parameter space. 
 
\begin{figure}[t!]
\centering
\includegraphics[width=10cm]{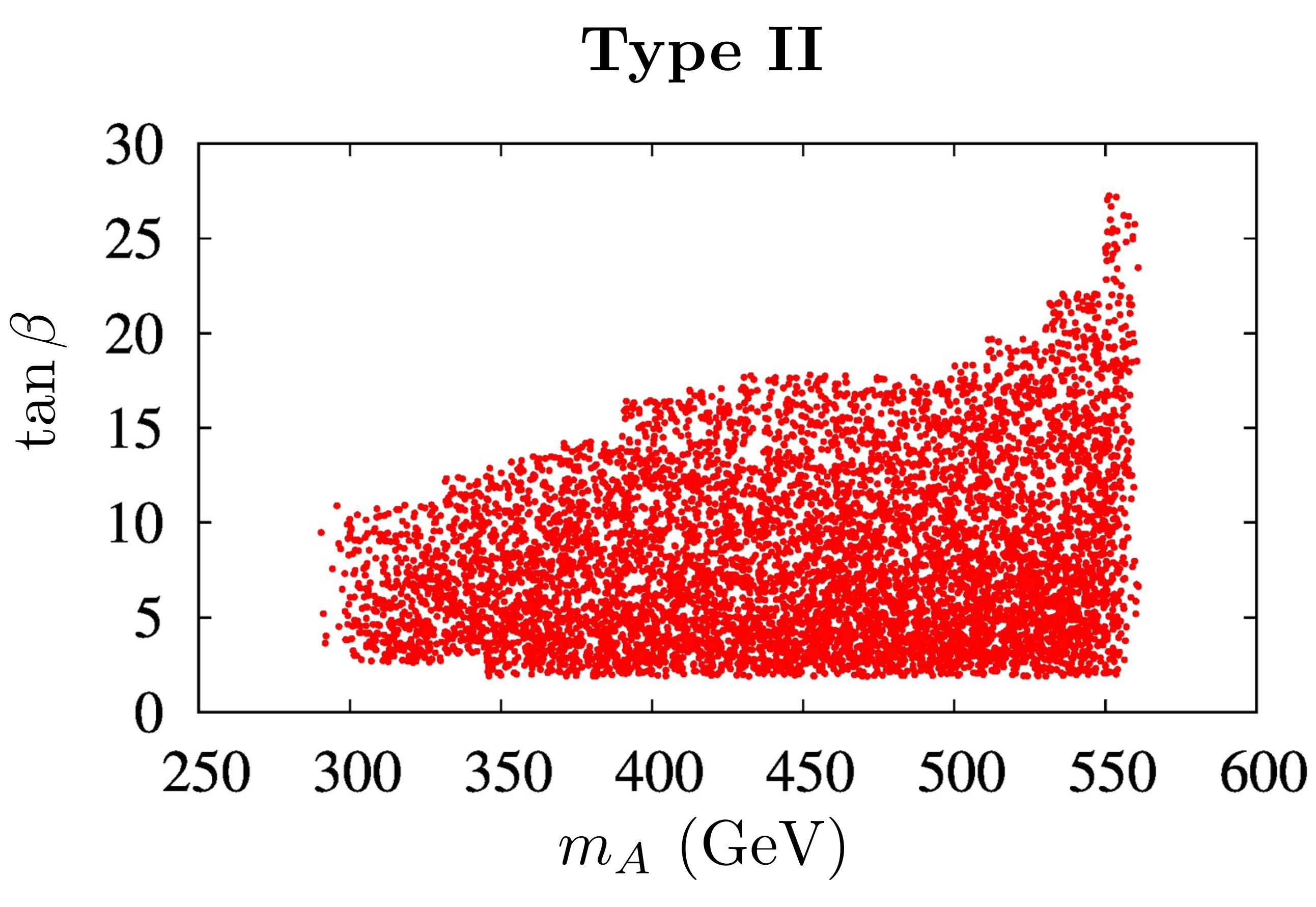}
\caption{\label{fig:type2tbmA}
Allowed parameters by the LHC Higgs data in the plane $(m_A,\tb)$ for Type II.
}
\end{figure}
 
\textit{(iii) Strong constraints from the} LHC \textit{heavy Higgs data in Type} II:
The 8 TeV LHC heavy Higgs searches in the decay channels of $\rr$, $\ttau$, $Z^0Z^0$,
and $W^+W^-$
start to constrain the 2HDM~\cite{Craig:2015jba}.
We find that especially Type II is very sensitive to the results
and thus does not allow too large $\tb$.
In Fig.~\ref{fig:type2tbmA},
we present the allowed parameters by the LHC Higgs data in the plane $(m_A,\tb)$
for Type II.
The heavy Higgs search results remove two regions,
small $\tb$ region with $m_A \lsim 2 m_t$
and large $\tb$ region. 

The small $\tb$ region is excluded by the $\rr$ channel 
through the gluon fusion production, similarly to the $h^0 =\hobs$ case~\cite{Song:2014lua}.
Both $g$-$g$-$A^0$ and $\gm$-$\gm$-$A^0$ vertices are loop induced,
mainly through the top quark loop.
Since all of four types have top quark Yukawa couplings with $A^0$
be inversely proportional to $\tb$,
small $\tb$ yields sizable gluon fusion production 
as well as sizable branching ratio into
$\rr$.
If $m_A = 330\gev$, for example,
$\tb$ should be above 2.5.
This is stronger than the constraints
from $b \to s \gm$ and 
$\Delta M_{B_d}$ on small $\tb$ like $\tb\gsim 1$
although the FCNC bound depends on the charged Higgs boson mass.
Note that the exclusion of this small $\tb$ region is common for all four types.
As $m_A$ goes beyond the $\ttop$ threshold,
the main decay mode of $A^0$ is into $\ttop$,
with the branching ratio practically one unless $\tb$ is too large.

In Type II, large $\tb$ region 
is excluded mainly by $\bb\to A^0 \to \ttau$~\cite{Craig:2015jba}.
Here, both $b$ quark and $\tau$ Yukawa couplings with $A^0$ 
are proportional to $\tb$,
yielding the signal rate proportional to $\tb^4$.
A large portion of the parameter space is excluded.
For example, the $m_A =300\gev$ case excludes $\tb \gsim 10$. 
In Type X and Type Y,
the multiplication of $\tau$ and $b$ quark Yukawa couplings with $A^0$
does not have $\tb$ dependence.
The LHC8 $\bb\to A^0 \to \ttau$ constraint for Types X and Y is not strong yet.
In Type I, all of the Yukawa couplings are suppressed by $\tb$,
which is weakly constrained.

\begin{figure}[t!]
\centering
\includegraphics[width=8cm]{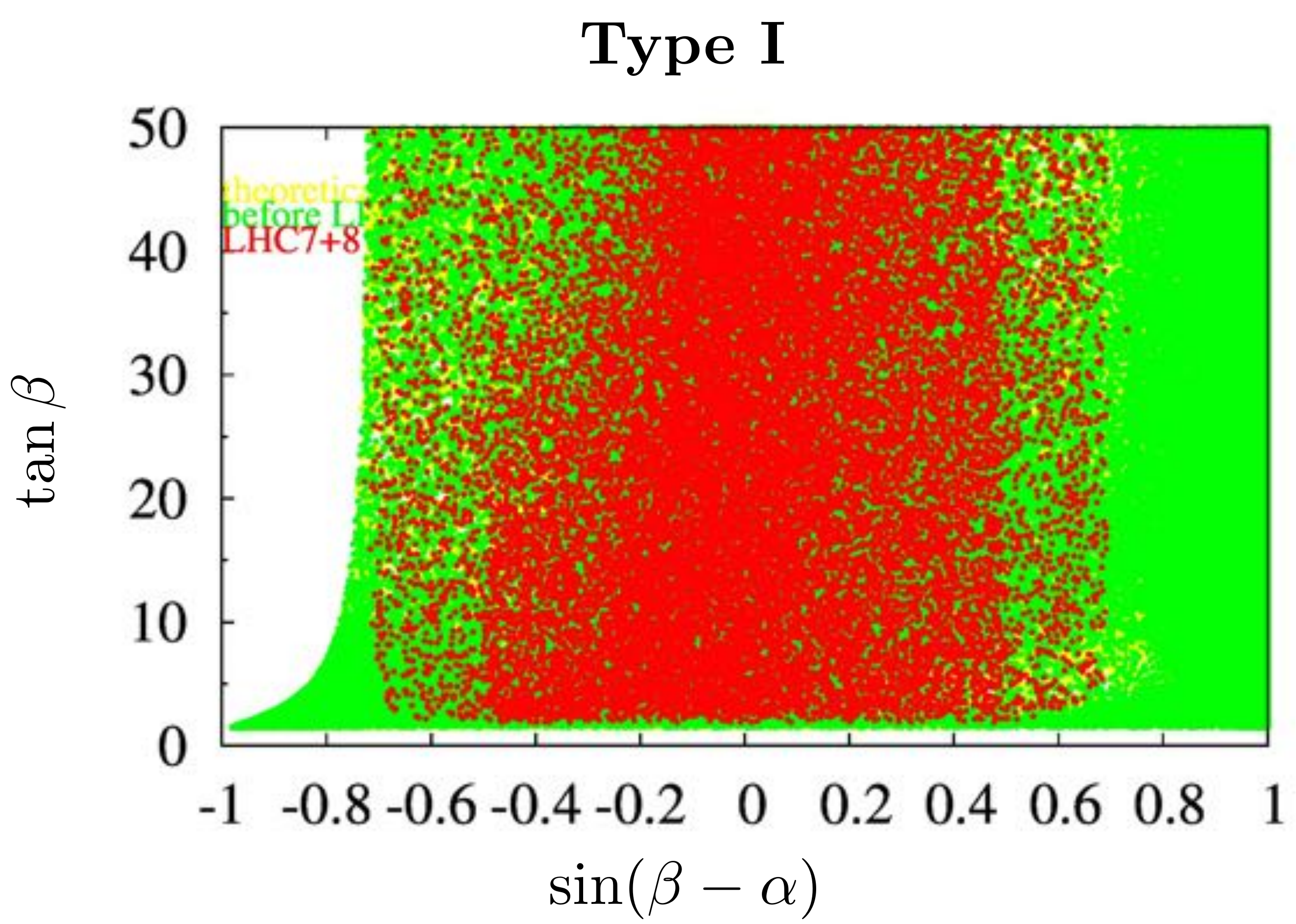}~~~~~
\includegraphics[width=8cm]{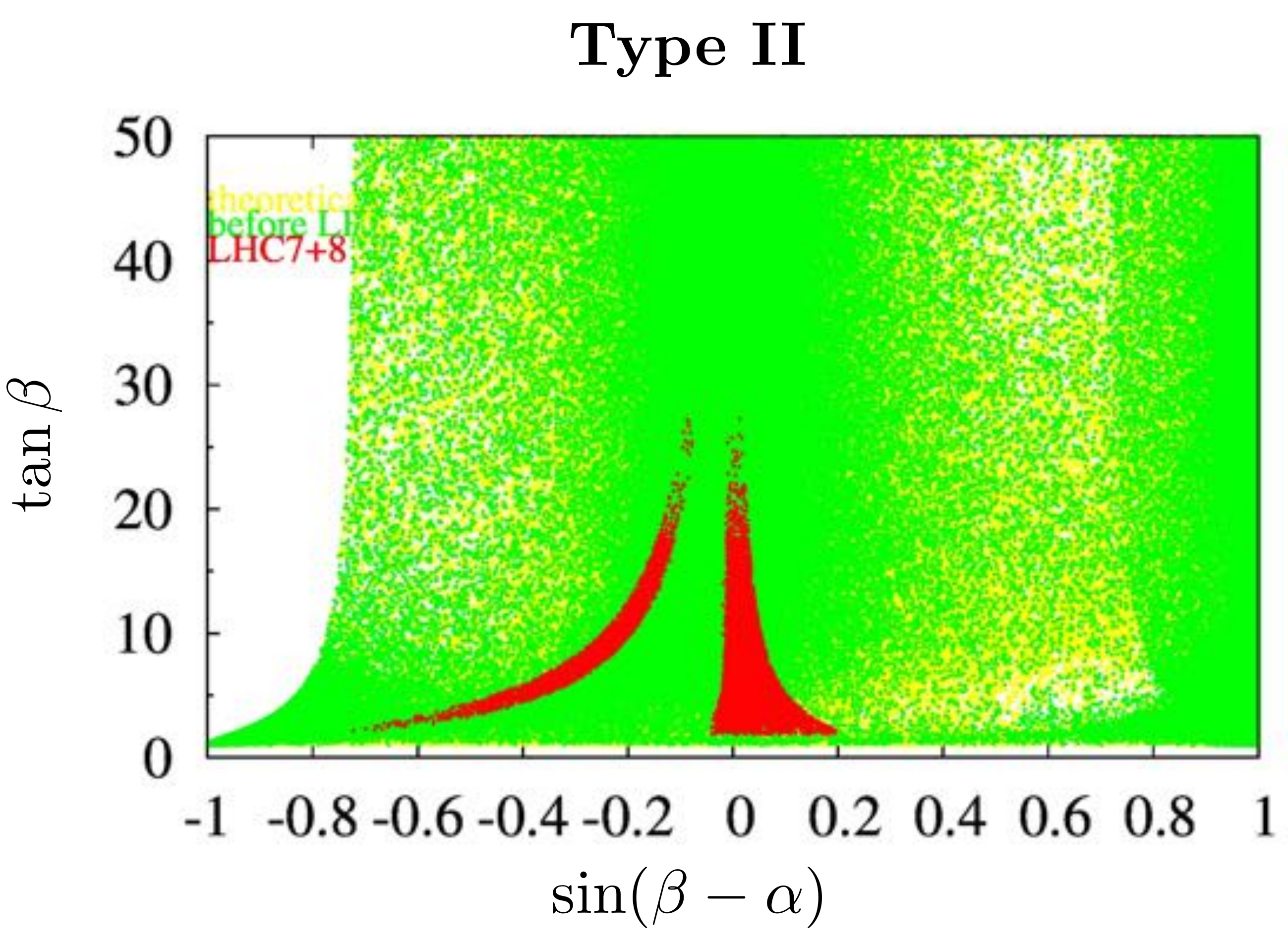}\\[0.3cm]
\includegraphics[width=8cm]{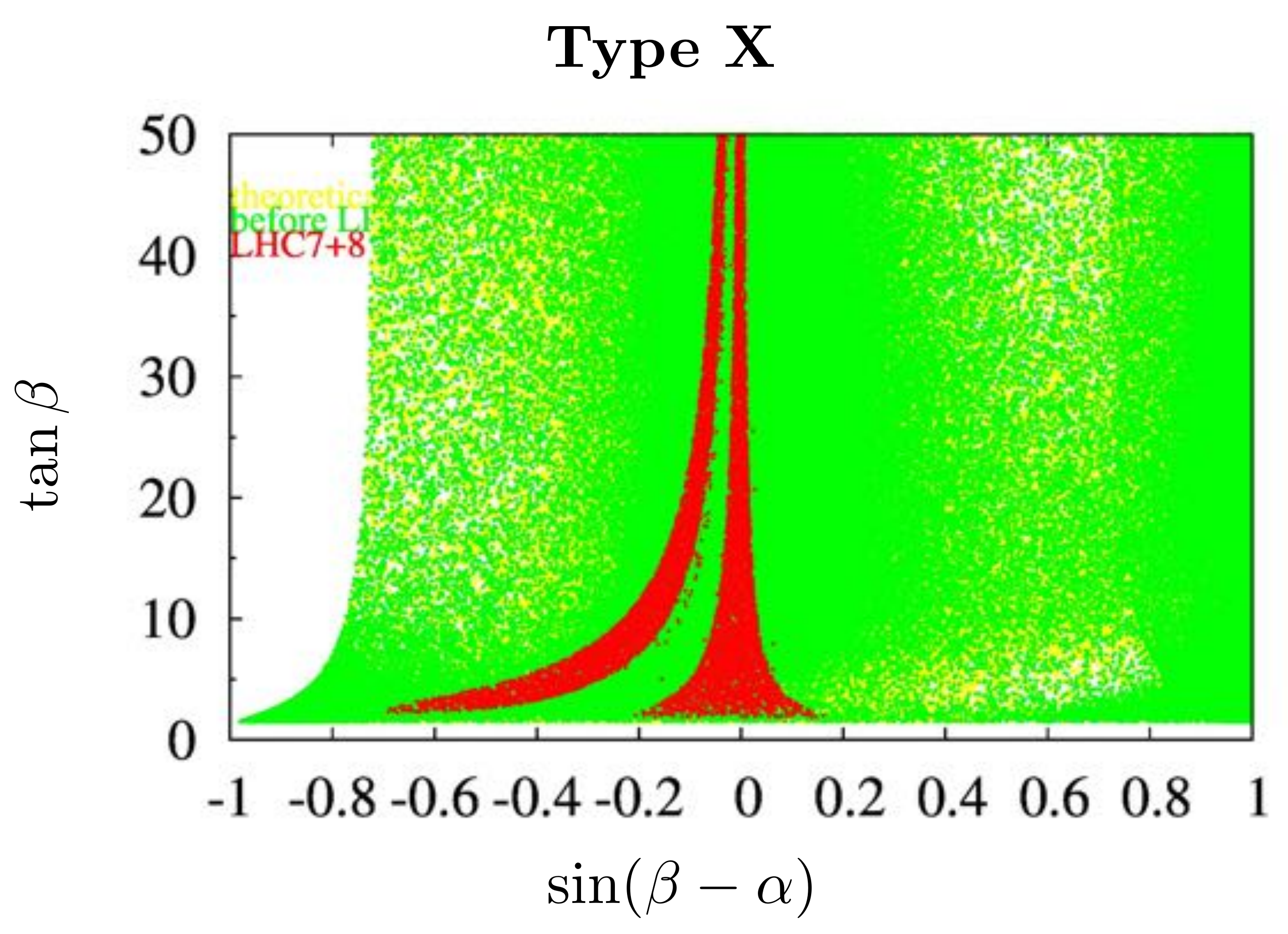}~~~~~
\includegraphics[width=8cm]{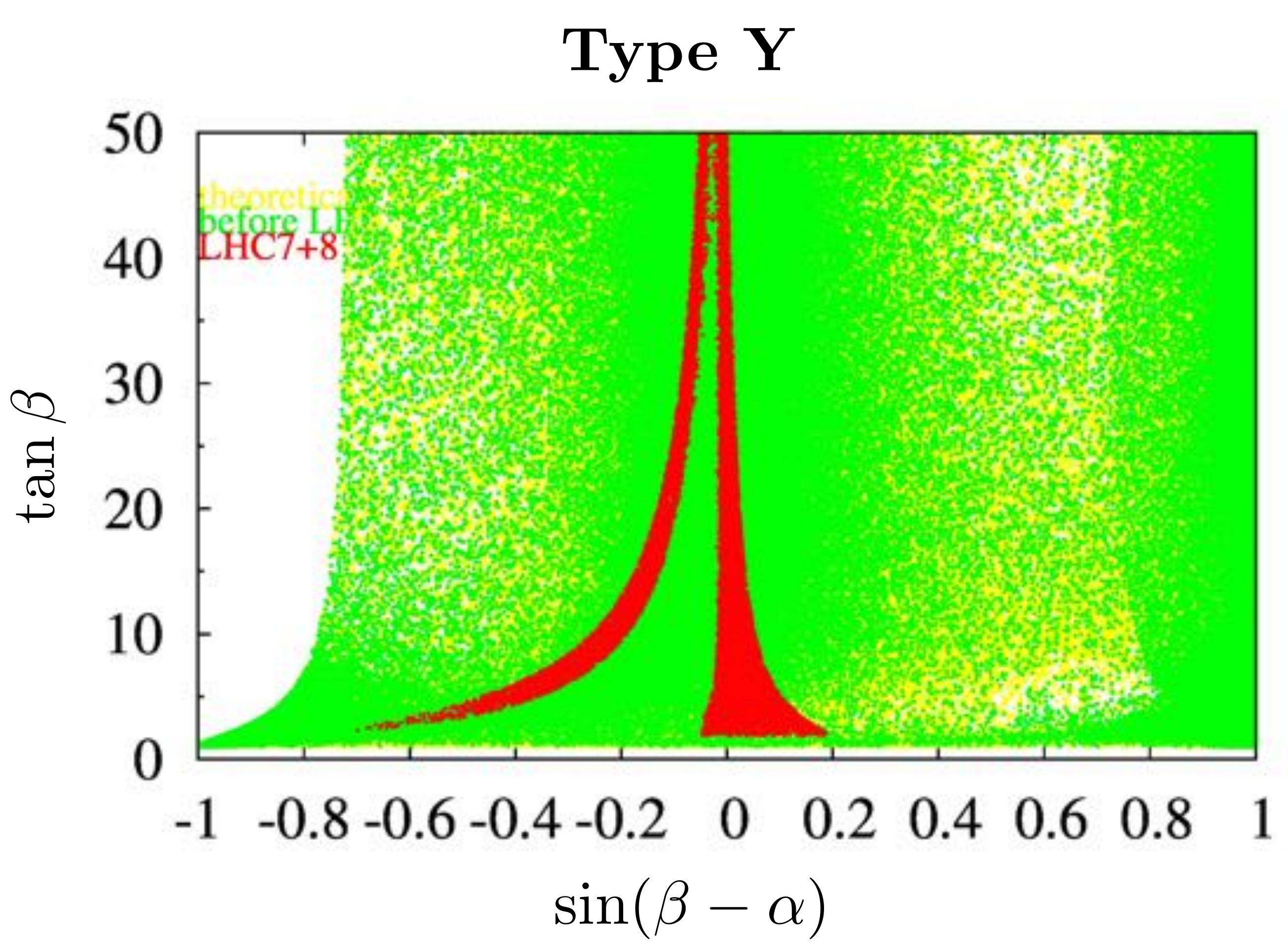}
\caption{\label{fig:tbsba}
Constraints in the $(\sin(\beta-\alpha),\tan\beta)$ plane for Type I, II, X, and Y.
Color scheme is the same as in Fig.~\ref{fig:mAmcH}.
 }
\end{figure}

\textit{(iv) Preferring the alignment limit}:
The deviation from the alignment limit is well parameterised by
$\sba$.
In Fig.~\ref{fig:tbsba},
we show the allowed parameter space in the $(\sba,\tb)$ plane
for Type I, II, X, and Y.
Type I allows sizable deviation: $|\sba|\lsim 0.5$ in most cases,
but even $|\sba|\simeq 0.7$ is allowed scarcely.
This is expected from the normalised Yukawa couplings in Eq.~(\ref{eq:Yukawa:couplings})
which have the $\tb$-suppressed deviation from the alignment.
The constraint from the LHC Higgs data on $\sba$ is relatively weak.
For other types, the $\tb$-enhanced Yukawa couplings constrain the model,
resulting in the preference to the alignment limit,
stronger for large $\tb$.

For Type II, X, and Y,
there are two separate regions of allowed  parameters.
One region is along the alignment line ($\sba=0$).
For small $\tb$
which weakens the $\tb$-enhanced deviations in Eq.~(\ref{eq:Yukawa:couplings}), 
$\sba$ can be as large as 0.2.
The other type of the allowed regions
is a bit apart from the $\sba=0$ line but toward negative $\sba$.
For small $\tb$, $\sba$ can be as large as $-0.6$.
As shown in Fig.~\ref{fig:type2tbmA},
Type II does not have the allowed region for $\tb> 20$ for the most part.

\begin{figure}[t!]
\centering
\includegraphics[width=10cm]{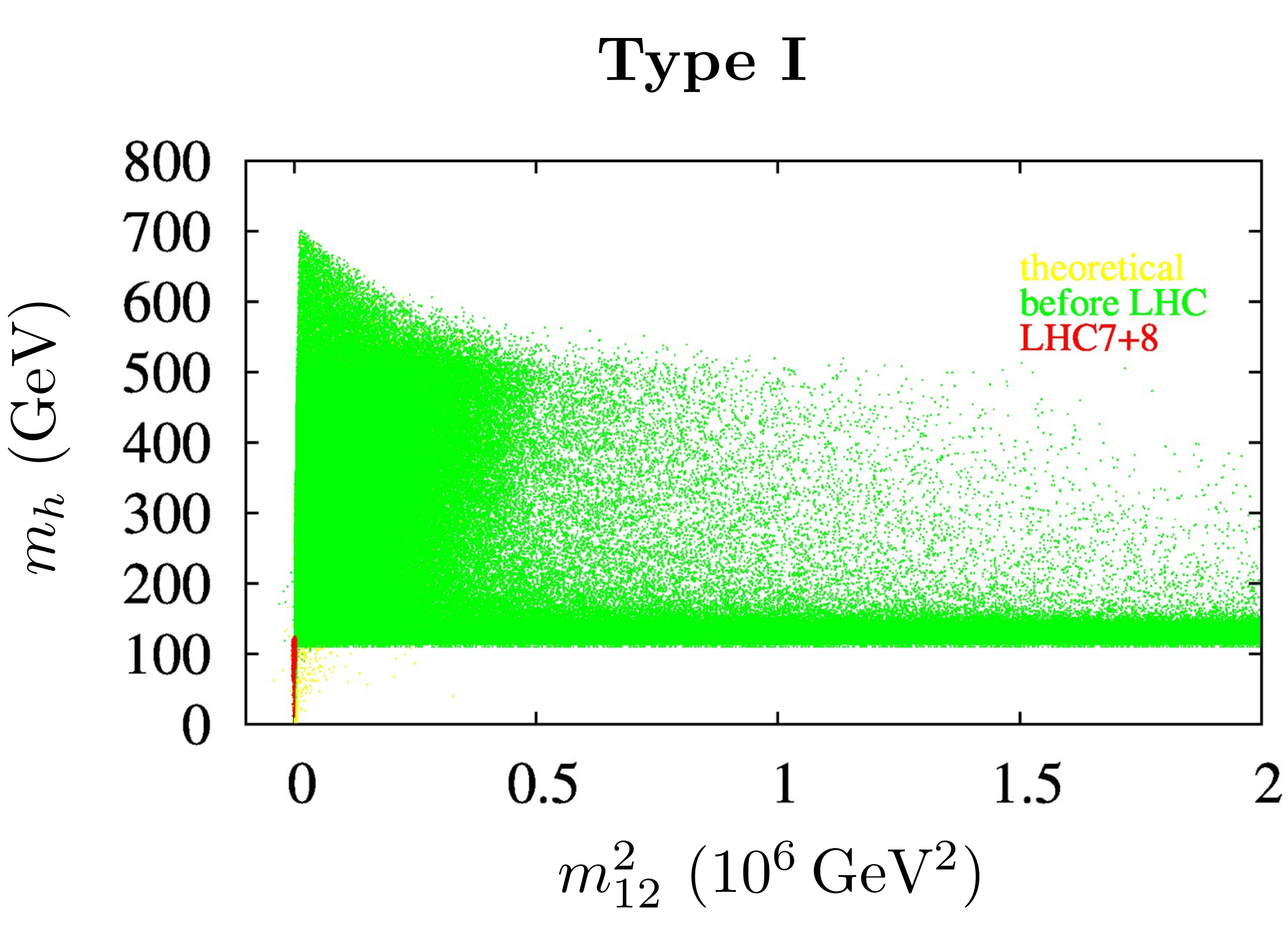}
\caption{\label{fig:type1mhm12sq_overall}
Constraints in the $(m_{12}^2, m_h)$ plane for Type I.
Color scheme is the same as in Fig.~\ref{fig:mAmcH}. }
\end{figure}

\textit{(v) Low scale for $m_{12}^2$}:
The $m_{12}^2$ term in the Higgs potential
breaks the $Z_2$ symmetry softly.
The explicit soft $Z_2$ breaking 
is to be understood as the result of spontaneous symmetry breaking 
in a more fundamental theory.
Without a knowledge of the origin and dynamics of the $Z_2$ breaking,
there is no guideline about the $m_{12}^2$ scale.
In the hidden light Higgs scenario, however,
the condition of $m_H = 125\gev$ constrains the $m_{12}^2$ scale
very strongly.
It is clearly shown in Fig.~\ref{fig:type1mhm12sq_overall}
by the allowed parameters in the $(m_{12}^2,m_h)$ plane for Type I.
Before the LHC data, $m_{12}^2$ is not limited, while
the LEP bounds exclude most of the parameter space for $m_h < 114\gev$.
The current LHC Higgs data (red points) do prefer the low scale of $m_{12}^2$, 
which implies that the $Z_2$ parity is a good approximate symmetry
in the scenario.

\medskip
\begin{figure}[t]
\centering
\includegraphics[width=8cm]{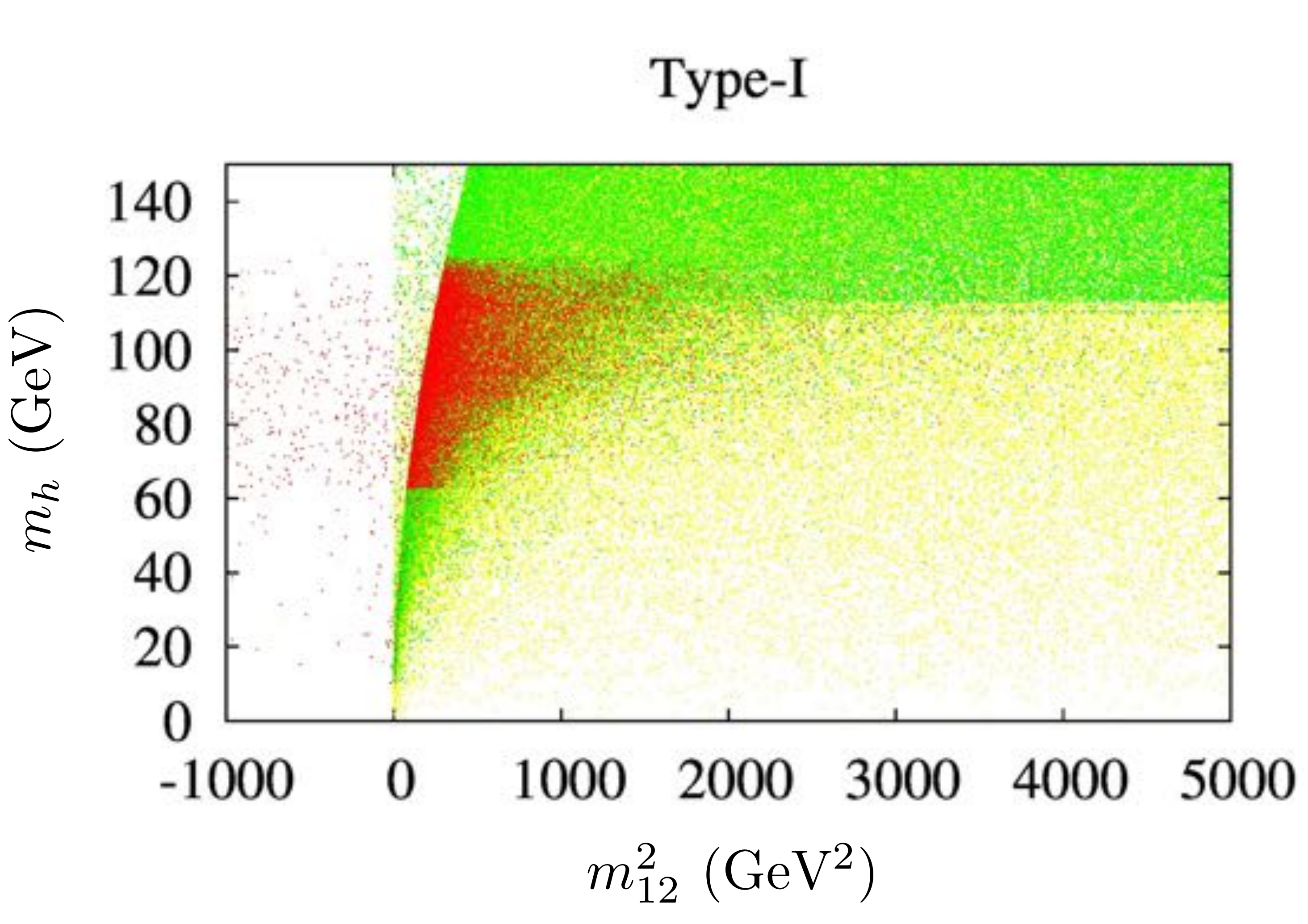}~~~~~
\includegraphics[width=8cm]{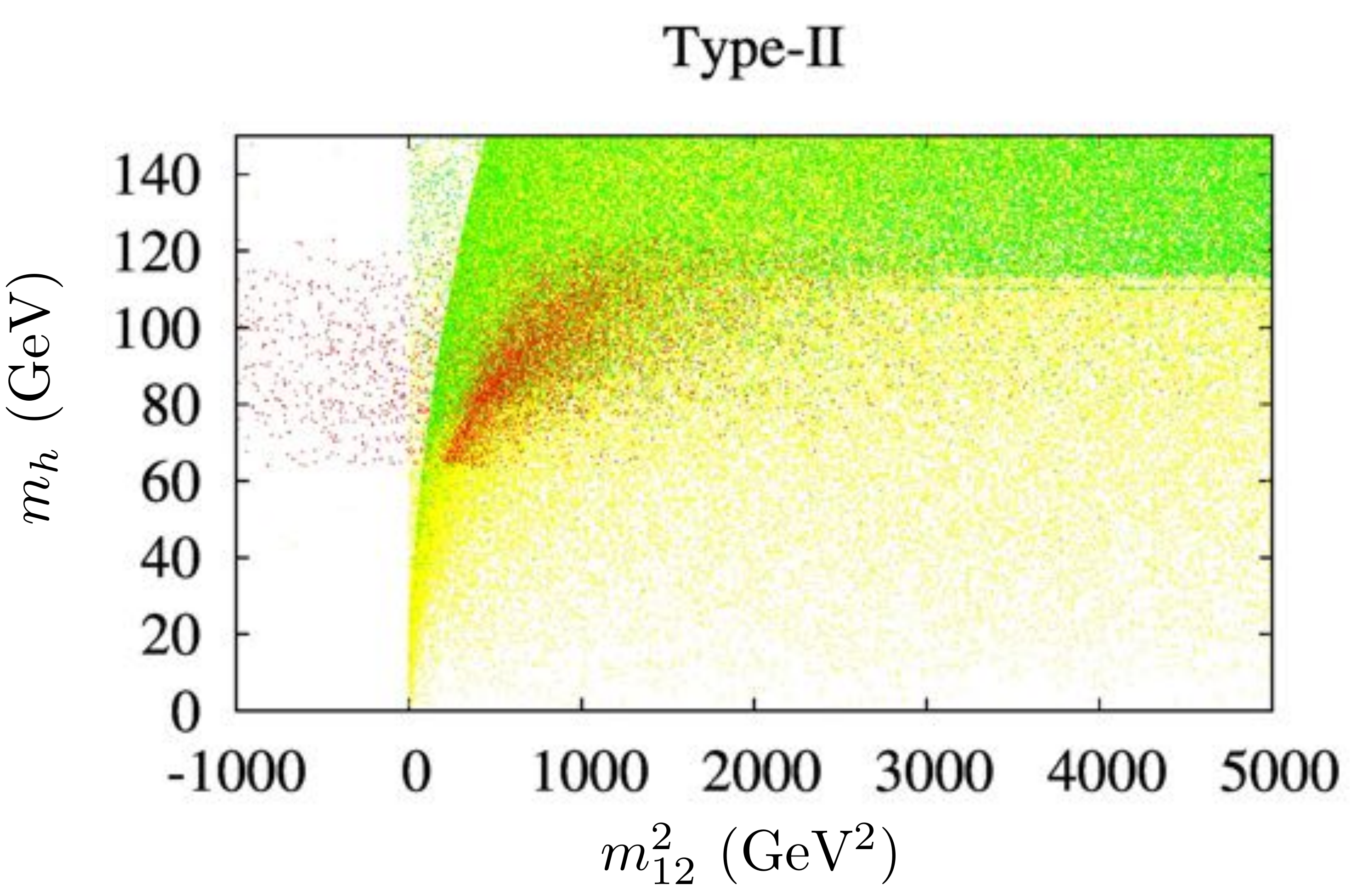}\\[0.3cm]
\includegraphics[width=8cm]{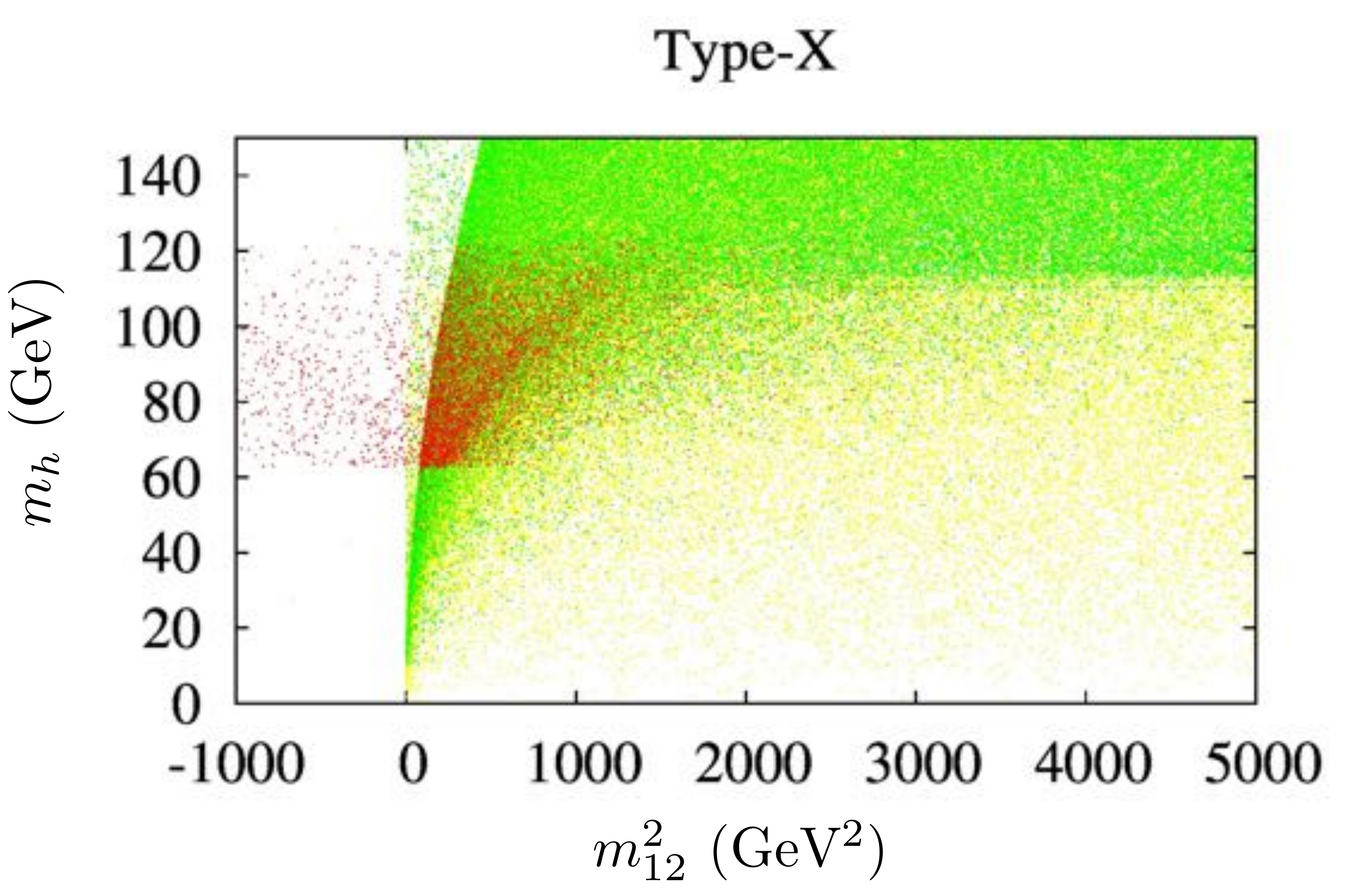}~~~~~
\includegraphics[width=8cm]{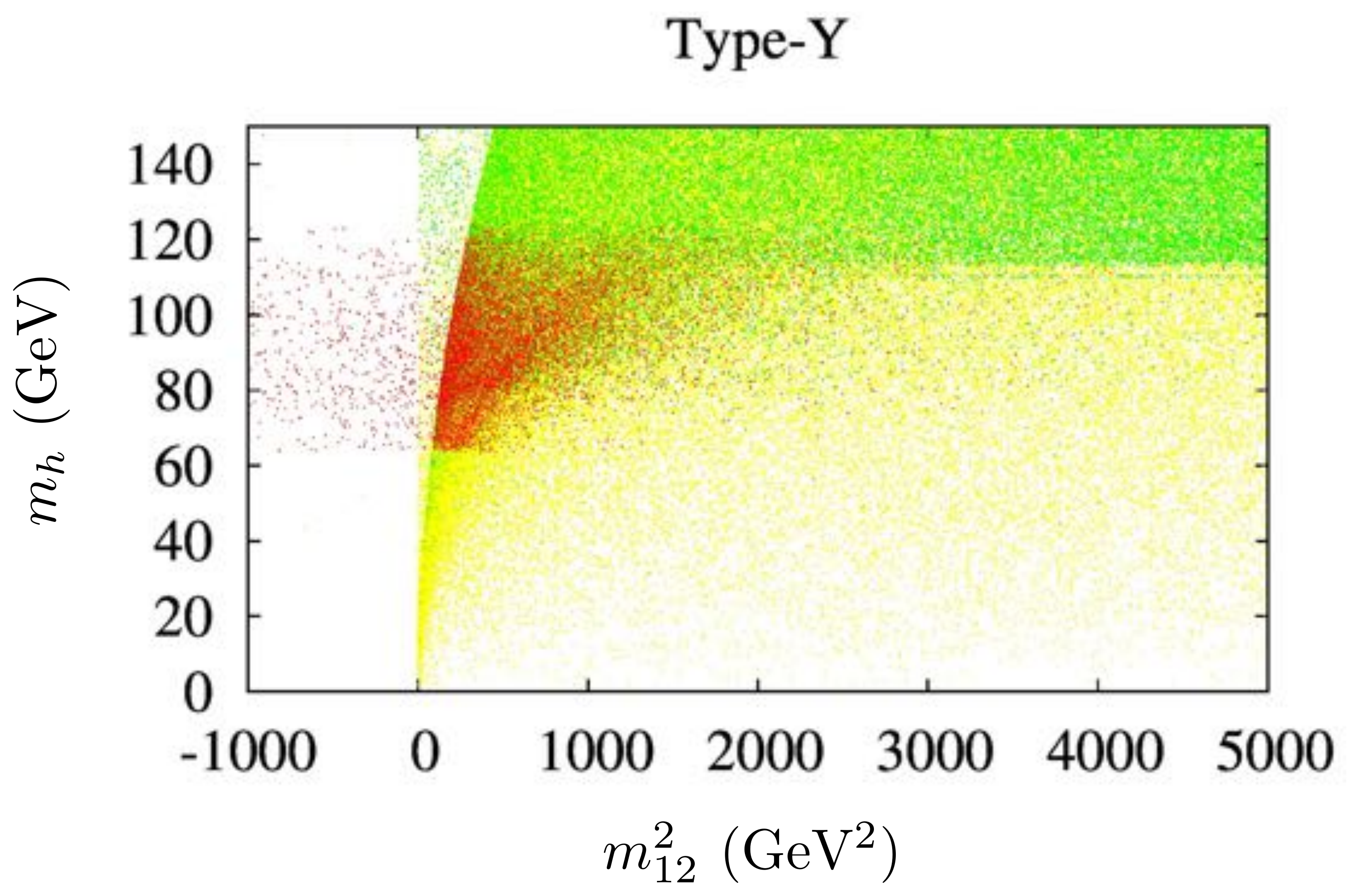}
\caption{\label{fig:mhm12sq}
Constraints in the $(m_{12}^2, m_h)$ plane for Type I and Type II.
Color scheme is the same as in Fig.~\ref{fig:mAmcH}.
 }
\end{figure}

In Fig.~\ref{fig:mhm12sq},
we show the constraints in the $(m_{12}^2, m_h)$ plane for Type I, II, X, and Y, 
focused on the LHC allowed regions.
All of the four types show similar shapes of the allowed regions:
one exception is Type II where the allowed space is much smaller.
As discussed before, this is because of the strong constraint from
the heavy Higgs boson search in the $b$ quark associated production
followed by the decay into $\ttau$.
Positive $m_{12}^2$ is preferred, although small negative value is also allowed.
The magnitude of $|m_{12}|$ is below about $45\gev$ mostly. 

Note that very light $h^0$ is possible in Type I.
However, too light scalar bosons may cause dangerous decay models such as $\eta \to \pi h^0$,
$\Upsilon \to A^0 \gm$, and $J/\Psi \to A^0 \gm$.
Although very light scalar bosons with mass below $10\gev$ 
are not excluded in some parameter space~\cite{Larios:2002ha,Anderson:2003bj,Clarke:2013aya},
we require for both $h^0$ and $A^0$
to be heavier than $10\gev$ for simplicity.
Another important constraint is from
the $H^0 \to h^0 h^0$ decay, which affects the LHC Higgs signal strength measurement.
In the alignment limit, the $H^0$-$h^0$-$h^0$ vertex, normalized by the SM
vertex $g_{hhh}^{\rm SM} = 3 m_{h_{\rm SM}}^2/v$,
is 
\bea
\label{eq:gHphiphi}
\gh_{Hh h} &=& \frac{1}{3}
\left[
1+2
\frac{m_h^2}{m_H^2}-2\left(\tb +\frac{1}{\tb}\right) \frac{m_{12}^2}{m_H^2}
\right] + \mathcal{O}(\sba).
\eea
In the alignment limit,
the decay $H^0 \to h^0 h^0$ 
is sizable in general.
In Types II, X, and Y where the alignment limit is strongly preferred 
(see Fig.~\ref{fig:tbsba}), 
$\gh_{Hh h}$ is too large to accommodate the LHC Higgs data
unless $m_{12}^2$ and $\tb$ are tuned to suppress $\gh_{Hhh}$.
It turns out that the LHC Higgs data prefer
$m_h> m_H/2$ in Type II, X, and Y in a general analysis 
based on the random generation of parameter points.
On the other hand, a very light Higgs boson with $m_h< m_H/2$ is allowed,
though less probable, when we carry out the analysis by choosing the special parameter choices leading to small enough $\hat{g}_{Hhh}$ as shown in Ref.~\cite{Bernon:2014nxa}.

\begin{figure}[t]
\centering
\includegraphics[width=8cm]{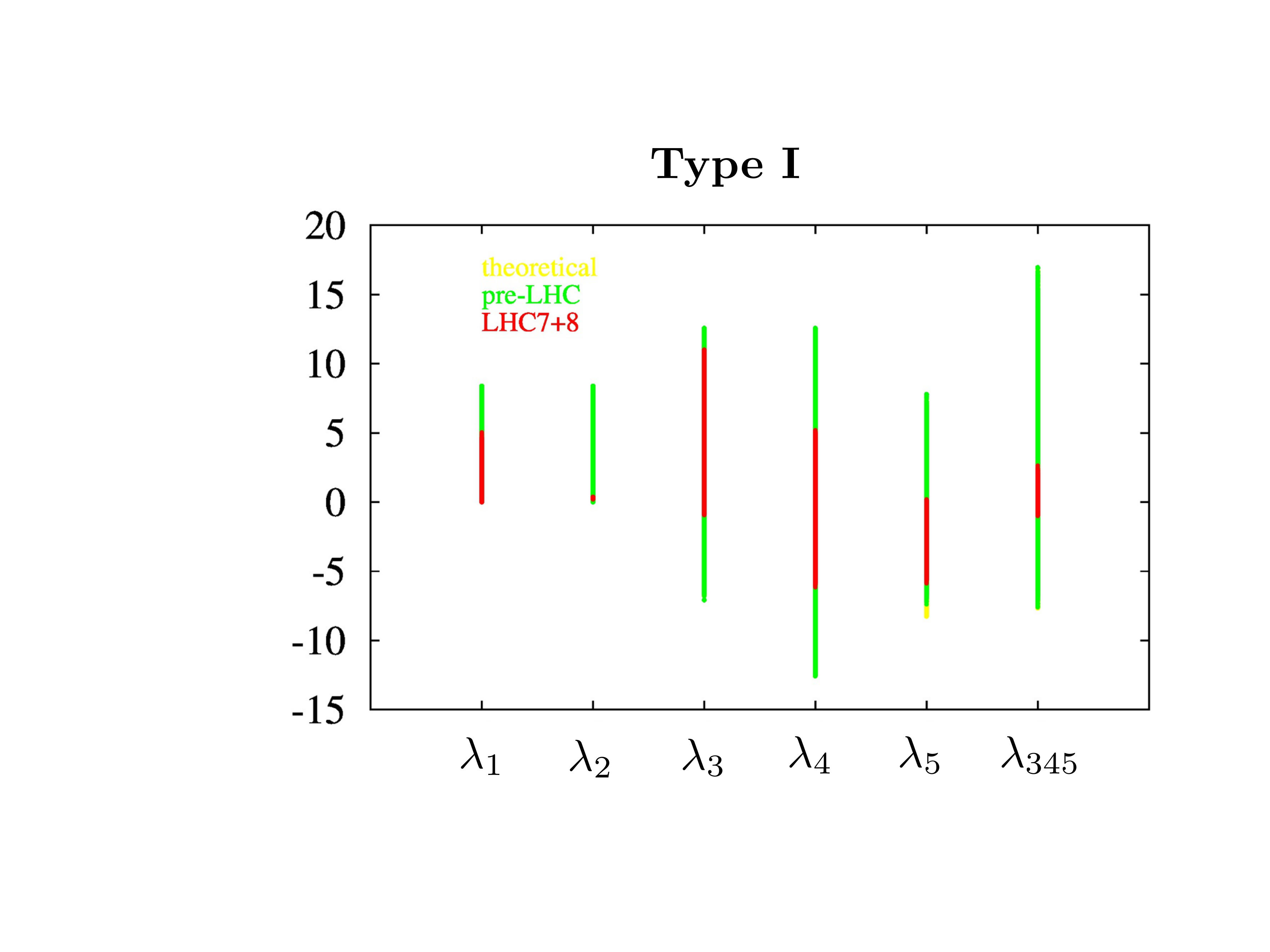}~~~~~
\includegraphics[width=8cm]{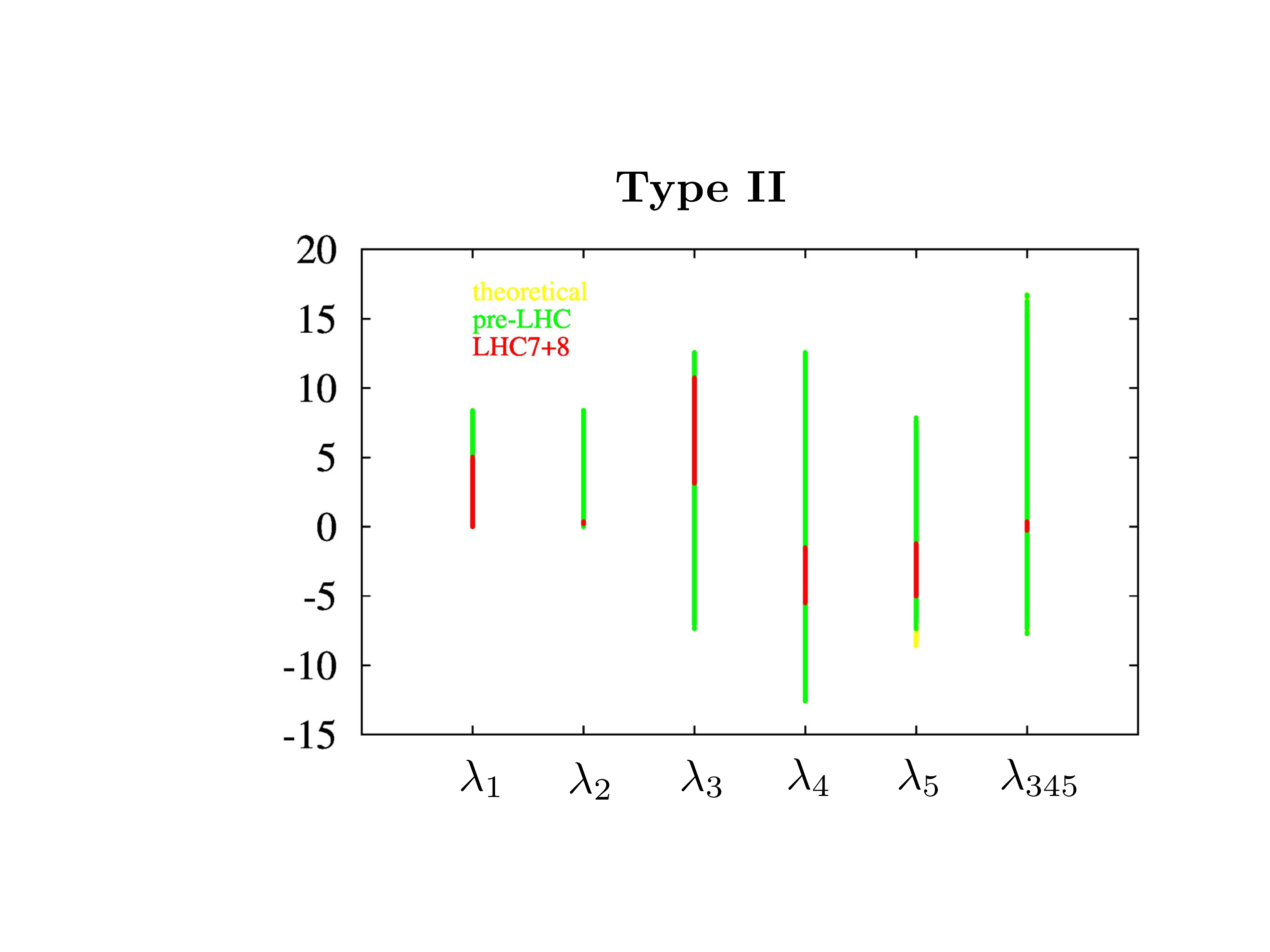}\\[0.3cm]
\includegraphics[width=8cm]{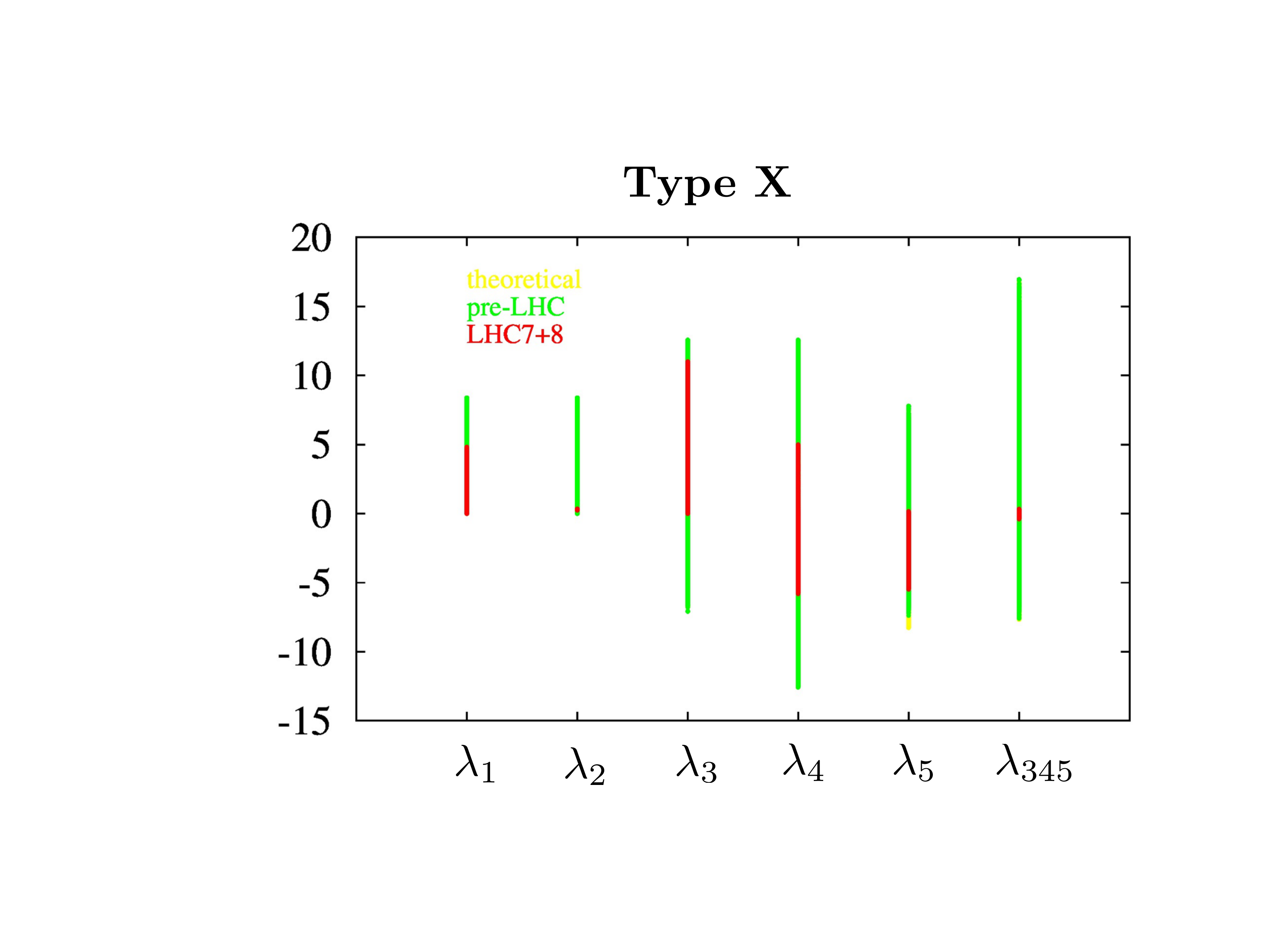}~~~~~
\includegraphics[width=8cm]{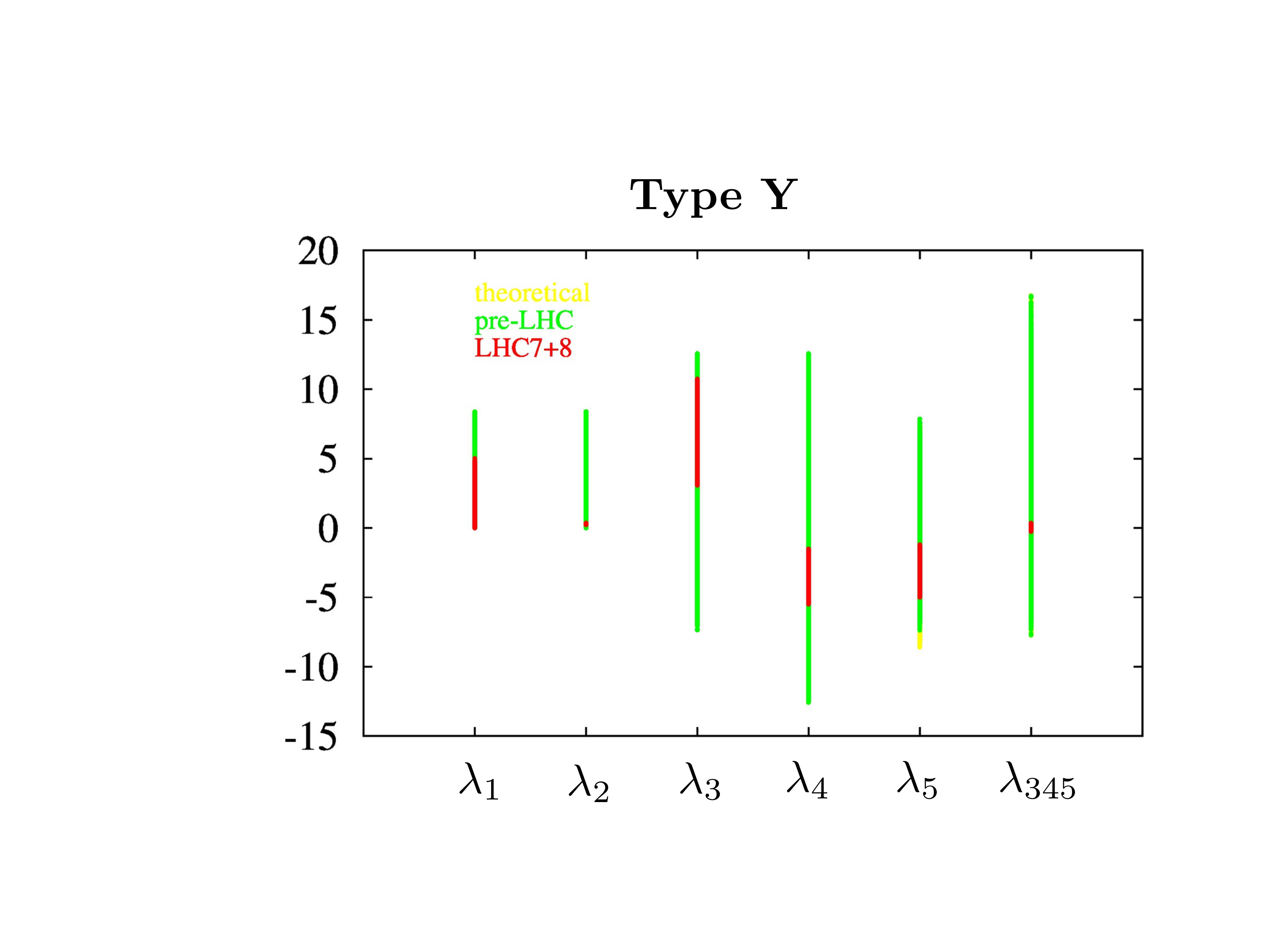}
\caption{\label{fig:lm}
Constraints on $\lm_{1,2,3,4,5}$ and $\lambda_{345}=\lambda_3+\lambda_4+\lambda_5$ 
for Type I, II, X, and Y.
Color scheme is the same as in Fig.~\ref{fig:mAmcH}.
 }
\end{figure}

\textit{(vi) Strong bound on $\lm_2$ but weak bounds on $\lm_{1,3,4,5}$}:
Figure \ref{fig:lm} shows how much the current LHC Higgs data constrain
the Higgs potential parameters $\lm_i$'s.
For all of four types,
$\lm_2$ is almost determined to be around $0.26$.
This is mainly by the mass measurement of $m_H = 125\gev$.
Since $m_{12}^2$ is small (see Fig.~\ref{fig:mhm12sq}) and $\tb>1$
from the FCNC constraints,
we have $m_H \sim \sqrt{\lm_2} v $ in the large $\tb$ limit: see Eqs.~(\ref{eq:Msqij})
and (\ref{eq:mhHmass}).
The condition of $m_H =125\gev$ almost fixes $\lm_2$.
Secondly, $\lm_{345}(\equiv \lm_3+\lm_4+\lm_5)$ is also limited by the observed Higgs boson mass 
in Types II, X, and Y.
The other $\lm_i$'s are not seriously constrained.
Compared to the pre-LHC (green) situation,
the LHC Higgs data reduce the value of $\lm_{1,3,4,5}$ by about half.
Considering the role of the current LHC Higgs signal strength measurement 
in determining $\lm_2$,
we anticipate that the values of $\lambda_{i}$ would be
substantially reduced if additional Higgs bosons are observed.

\section{Future prospect of $e^+ e^-\to Z^0 H^0 H^0$}
\label{sec:ee2ZHH}
\begin{figure}[t!]
\centering
\includegraphics[width=8cm]{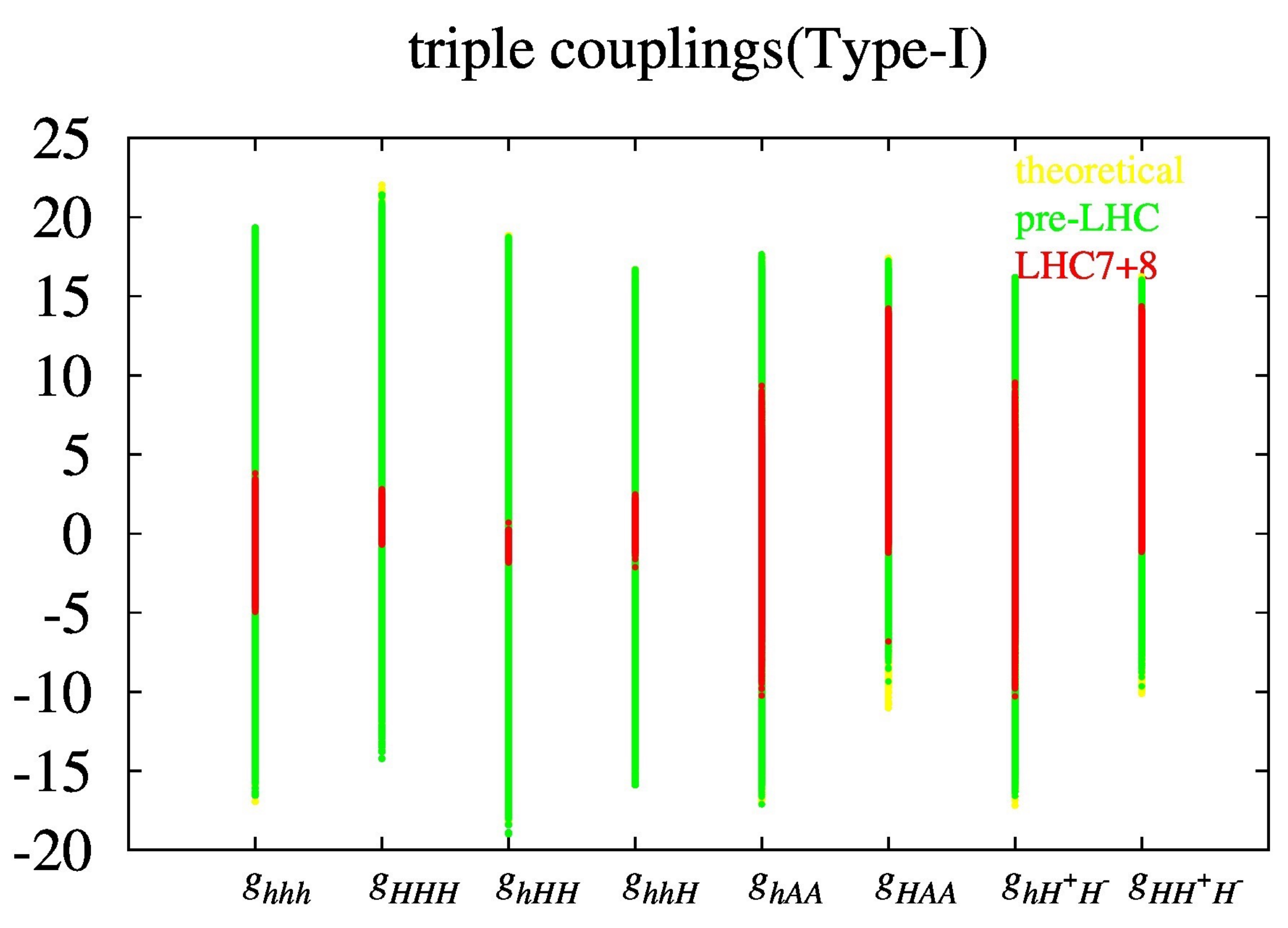}~~~~~
\includegraphics[width=8cm]{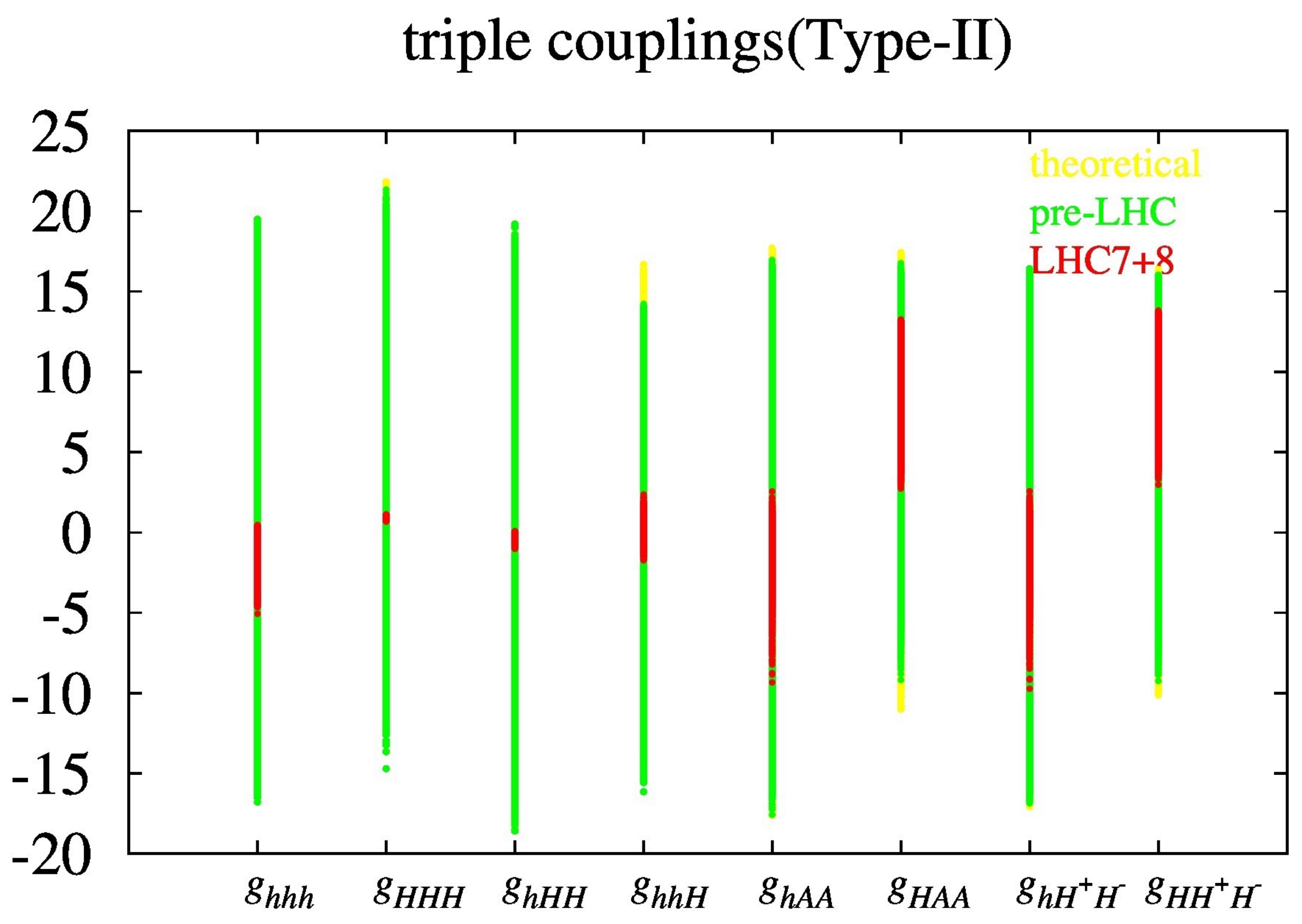}\\[0.3cm]
\includegraphics[width=8cm]{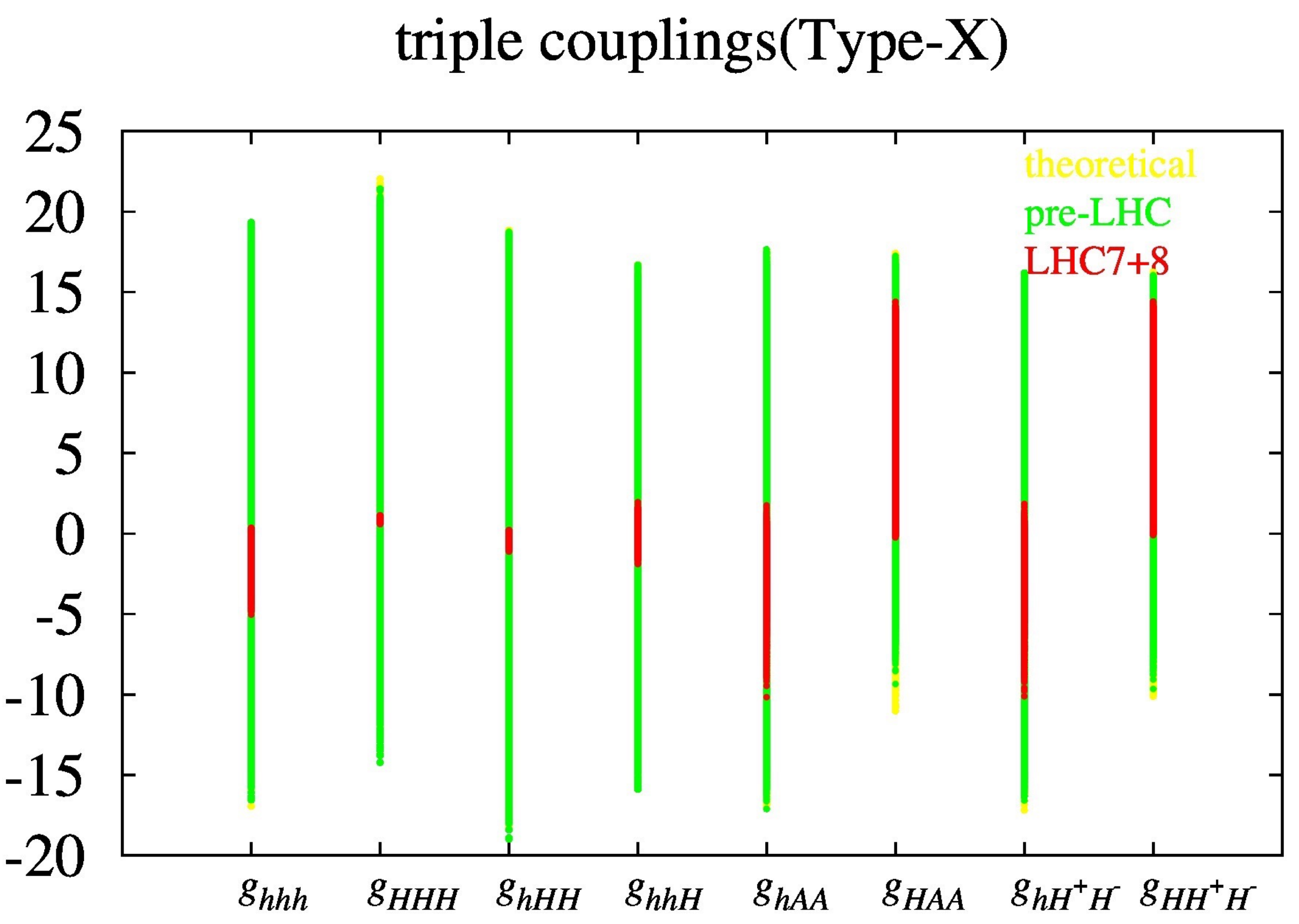}~~~~~
\includegraphics[width=8cm]{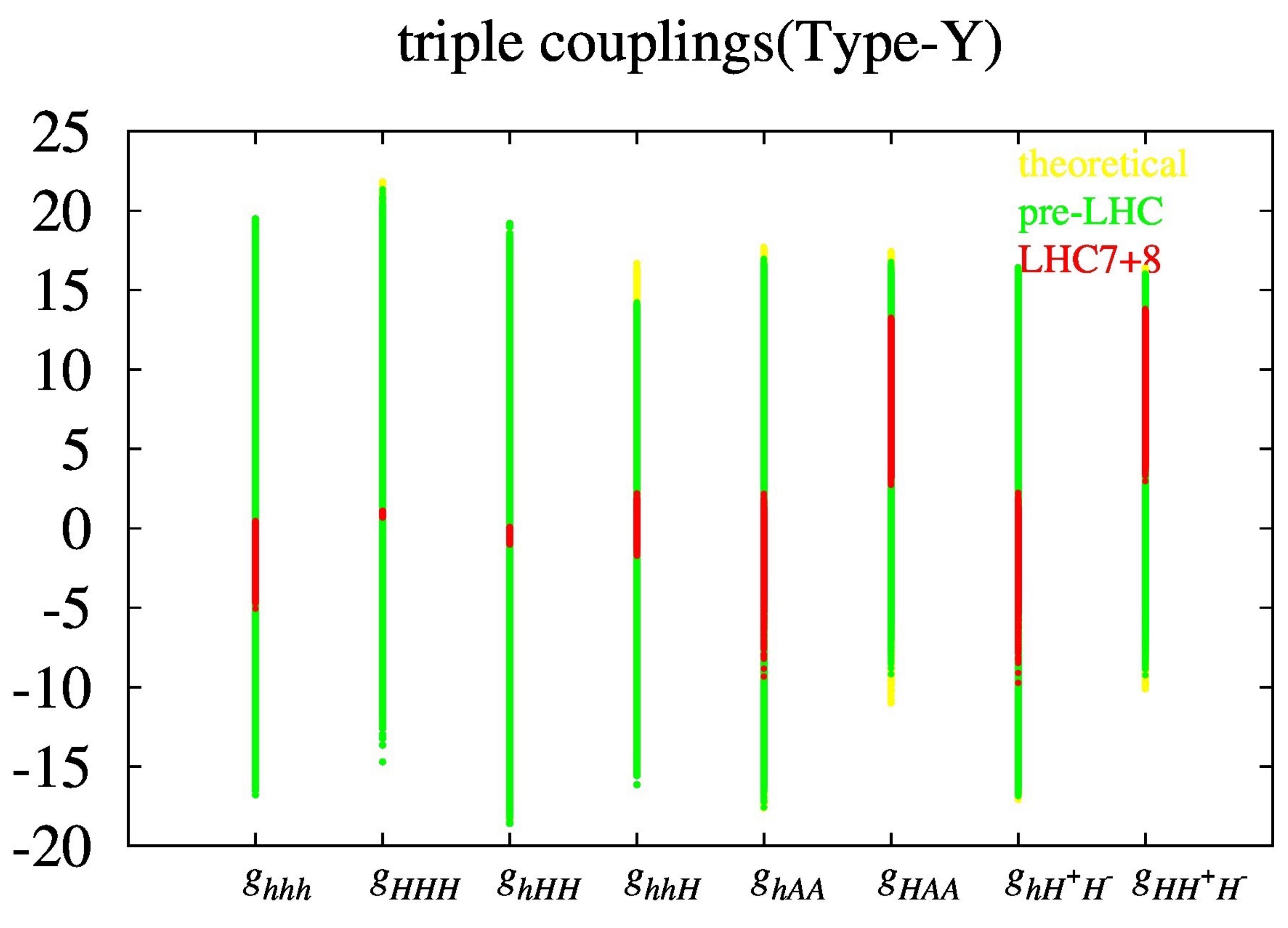}
\caption{\label{fig:triple}
Constraints on $\gh_{\phi_i \phi_j \phi_k}$ for Type I, II, X, and Y.
Color scheme is the same as in Fig.~\ref{fig:mAmcH}.
 }
\end{figure}

Focusing on the determination of the Higgs potential in the hidden light Higgs scenario,
we study the current status of the Higgs triple and quartic couplings.
First we study the allowed range of various Higgs triple couplings.
In the normal setup with $\hobs=h^0$,
Ref.~\cite{Baglio:2014nea} showed that the LHC Higgs data constrain
the normalised Higgs triple couplings $\gh_{hhh}$ lies between 0.56
and 1 at 95\% C.L. level in Type II. 
The Higgs boson pair production at a 14 TeV LHC 
was calculated for some benchmarks, not for the whole allowed parameter space,
since the physical properties of the heavy Higgs bosons are 
weakly constrained in the normal setup.
In the hidden light Higgs scenario,
the heavy Higgs boson properties
are shown to be significantly limited,
which may results in
higher predictability in the process involving the Higgs triple couplings.

In Fig.~\ref{fig:triple},
we present the allowed values of the normalised Higgs triple couplings, 
$\gh_{\phi_i \phi_j \phi_k}$ for $\phi_i = h^0, H^0, A^0, H^\pm$.
For all of four types, the triple coupling involving two 125 GeV states,
$\gh_{hHH}$ and $\gh_{HHH}$,
are quite strongly constrained.
Compared to the pre-LHC constraint,
the LHC data restrict $\gh_{hHH}$ and $\gh_{HHH}$ within a few percent.
In particular, $\gh_{HHH}$ in Types II, X, and Y
are very limited between 0.69 and 1.1.
In Type I that allows much larger parameter space,
the reduction is into 10\% level.

\begin{figure}[t!]
\centering
\includegraphics[width=8cm]{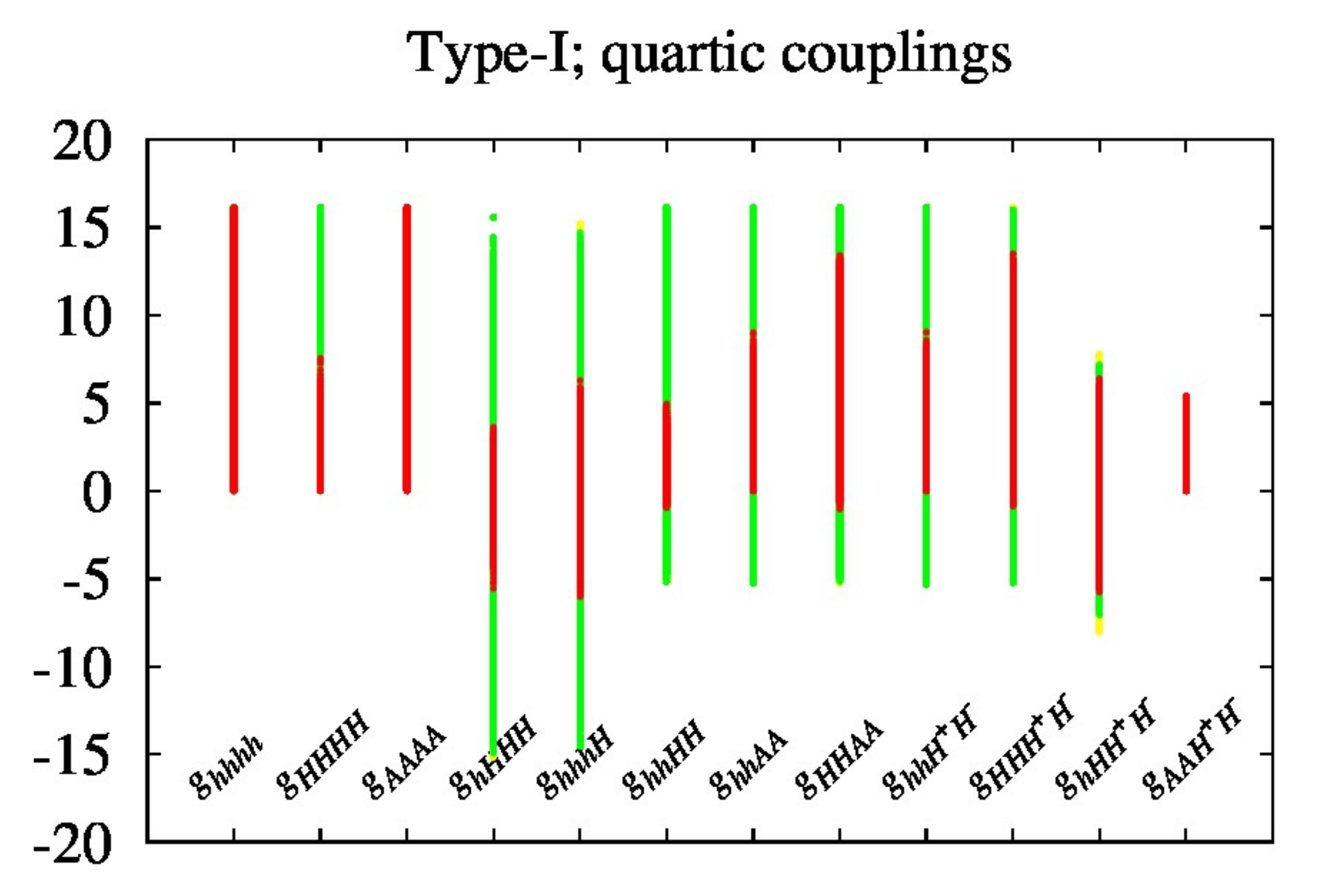}~~~~~
\includegraphics[width=8cm]{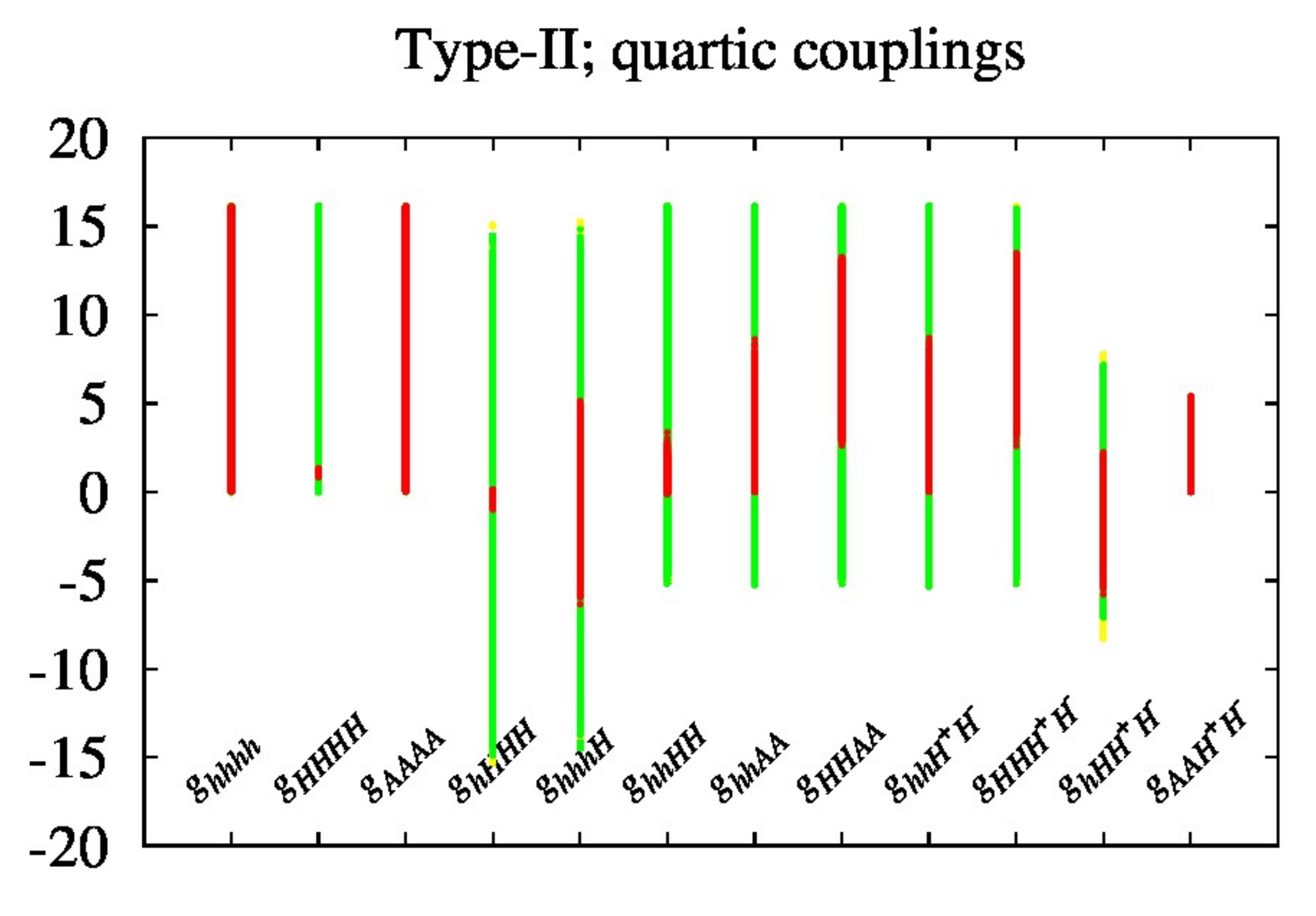}\\[0.3cm]
\includegraphics[width=8cm]{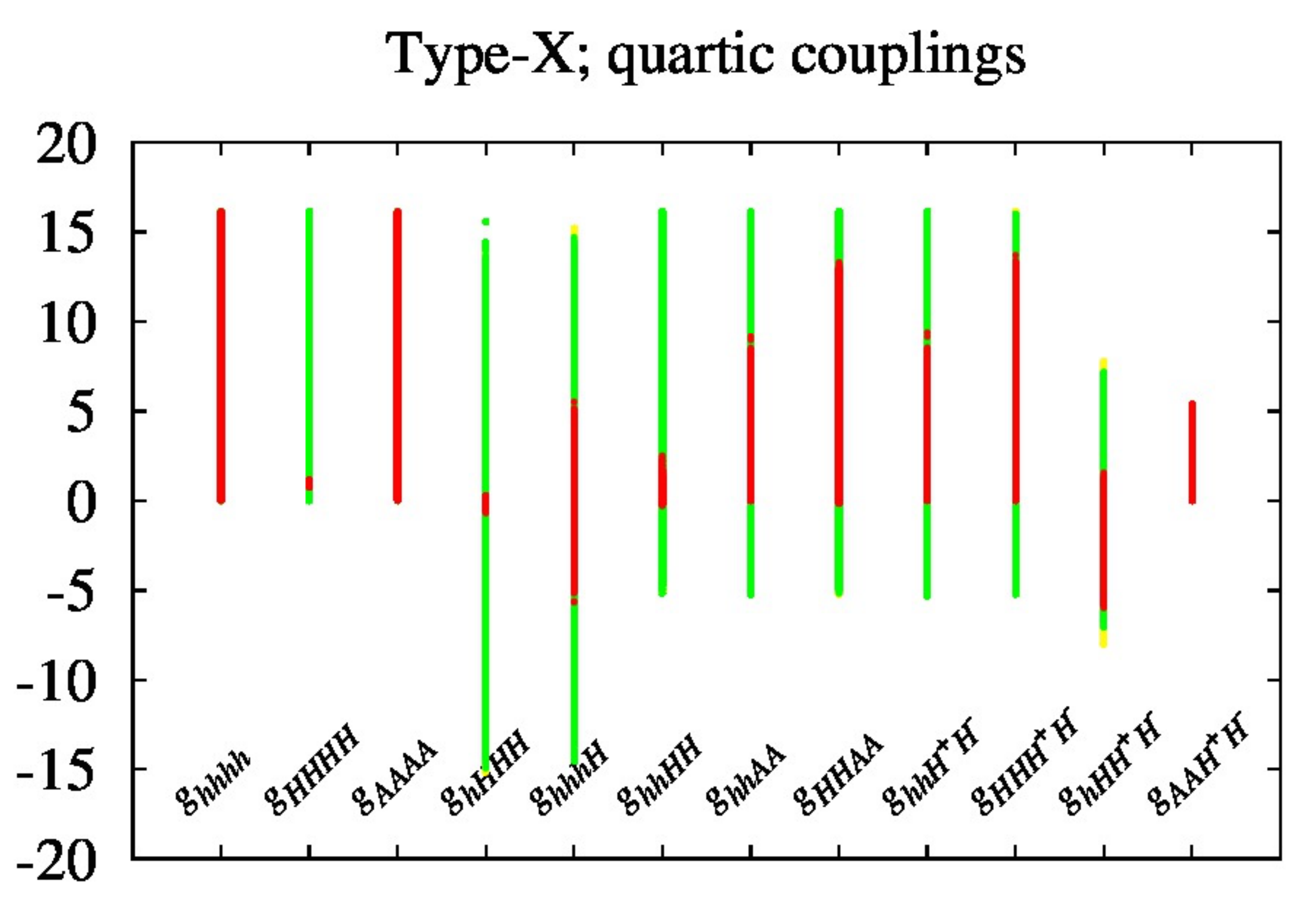}~~~~~
\includegraphics[width=8cm]{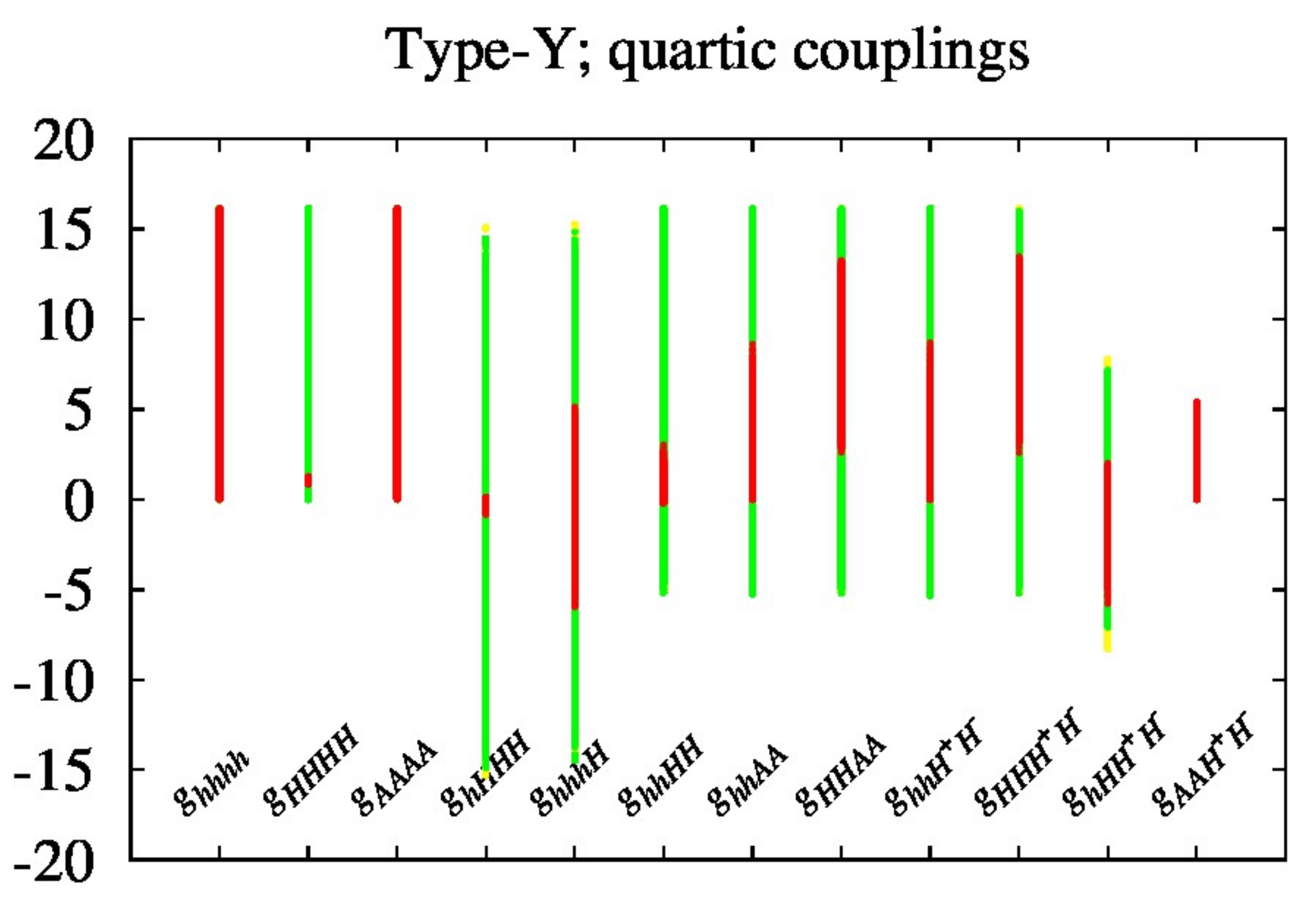}
\caption{\label{fig:quartic}
Constraints on $\gh_{\phi_i \phi_j \phi_k\phi_\ell}$ for Type I, II, X, and Y.
Color scheme is the same as in Fig.~\ref{fig:mAmcH}.
 }
\end{figure}

The quartic Higgs couplings are in general less constrained
even in the hidden light Higgs scenario.
In Fig.~\ref{fig:quartic},
we show the normalized quartic couplings by the SM value, 
with the three step constraints imposed.
It is remarkable that $\gh_{HHHH}$ and $\gh_{hHHH}$
in Types II, X, and Y
are quite significantly constrained as to be similar to the SM value.
On the contrary, Type I does not have limited value of $\gh_{HHHH}$ and $\gh_{hHHH}$
because of the large allowed parameter space.
Other quartic Higgs couplings can be very large compared with the SM value.
Contrary to the SM case where the quartic coupling will remain  
unaccessible due to the tiny cross section of 
$e^+ e^- \to Z^0 \hsm\hsm\hsm$~\cite{Battaglia:2001nn,Beyer:2006hx},
some large quartic couplings
can yield large enough cross section.
In all of the four types,
$\gh_{hhhh}$ and $\gh_{AAAA}$
can have the enhancement factor more than ten,
which can be probed, \textit{e,g,}, through $e^+ e^- \to Z^0 h^0 h^0 h^0$
and $\rr\to A^0 A^0 A^0$, respectively.

\begin{figure}[t!]
\centering
\includegraphics[width=8cm]{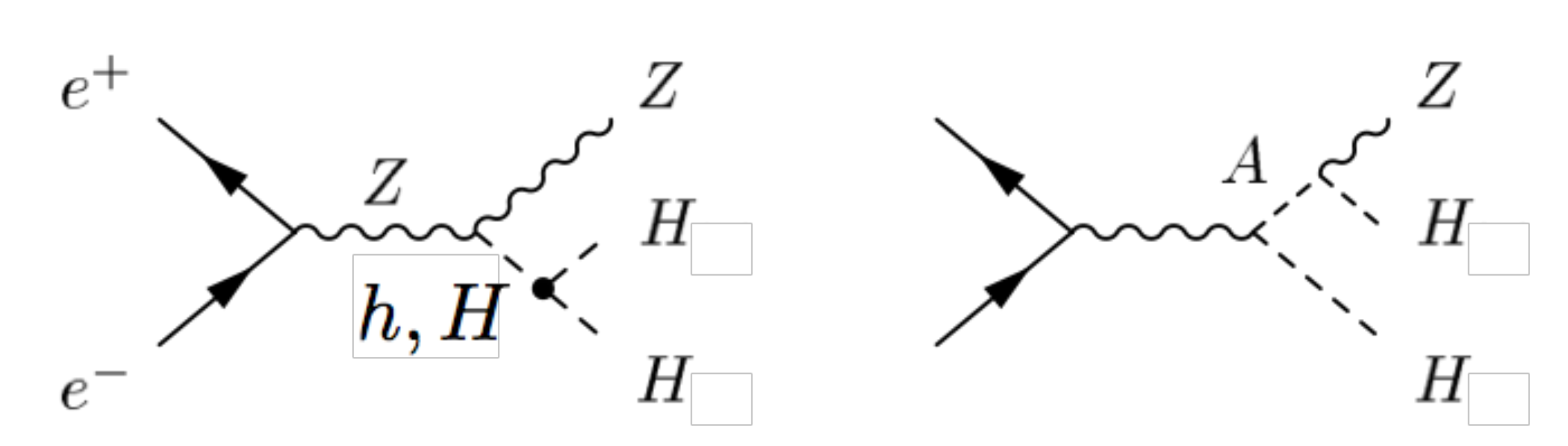}~~~~~
\includegraphics[width=8cm]{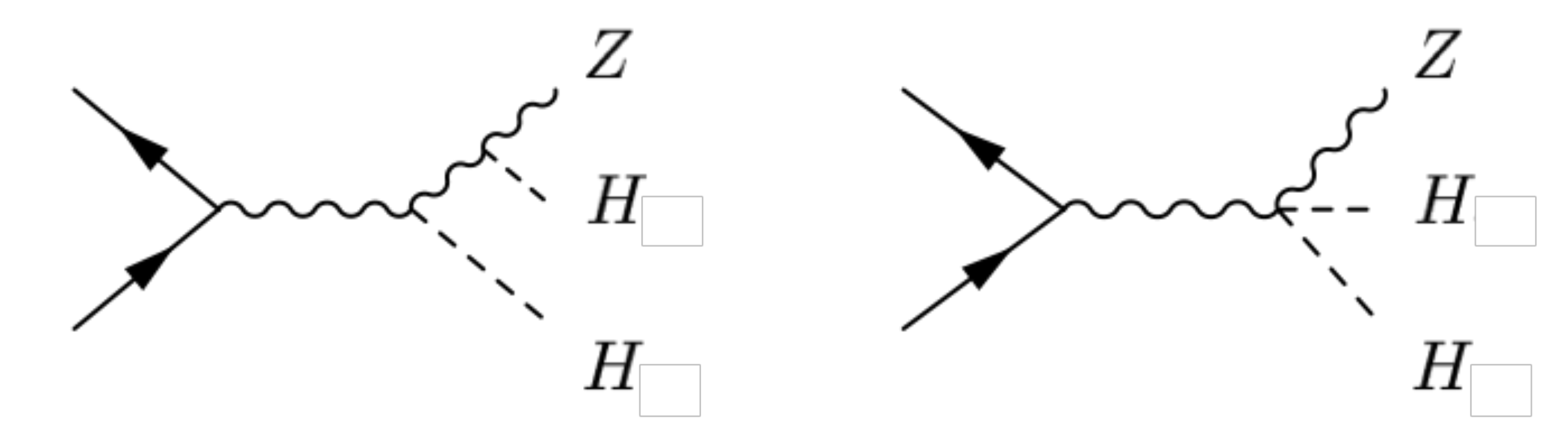}
\caption{\label{fig:eeZHH:feyn}
Feynman diagrams for $e^+ e^- \to Z^0 H^0 H^0$.
 }
\end{figure}

In order to probe $\gh_{hHH}$ and $\gh_{HHH}$,
we study the double Higgs-strahlung at an $e^+ e^-$ collider,
$e^+ e^- \to Z^0 H^0 H^0$~\cite{Djouadi:2005gj,Kilian:1995tr,Djouadi:1999gv}.
The Feynman diagrams are presented in Fig.~\ref{fig:eeZHH:feyn}.
Since the physical properties of $H^0$ are well known,
the Higgs boson can be used as a tagging particle for a new physics model.
$\gh_{hHH}$ and $\gh_{HHH}$ contribute in the first diagram.
The second Feynman diagram shows that the $Z^0$-$H^0$-$A^0$ vertex 
also contributes.
As classified in Eq.~(\ref{eq:sba:cba}),
the $Z^0$-$H^0$-$A^0$ vertex is proportional to $\sba$,
which is suppressed in the alignment limit.
If $\sba\neq 0$ and the kinematical space includes the pole of the $A^0$ propagator,
the total cross section can be highly enhanced.

\begin{figure}[t!]
\centering
\includegraphics[width=8cm]{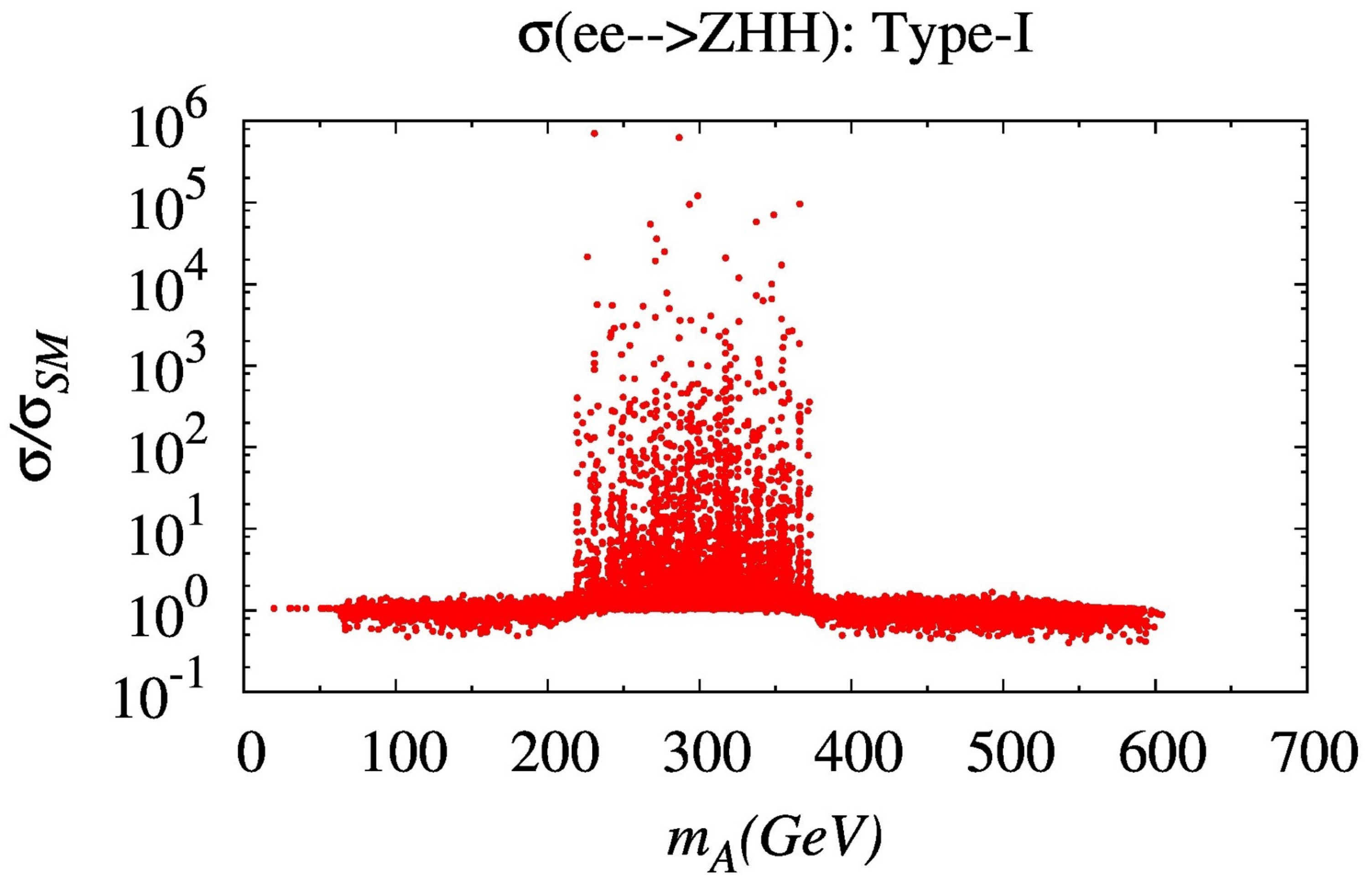}~~~~~
\includegraphics[width=8cm]{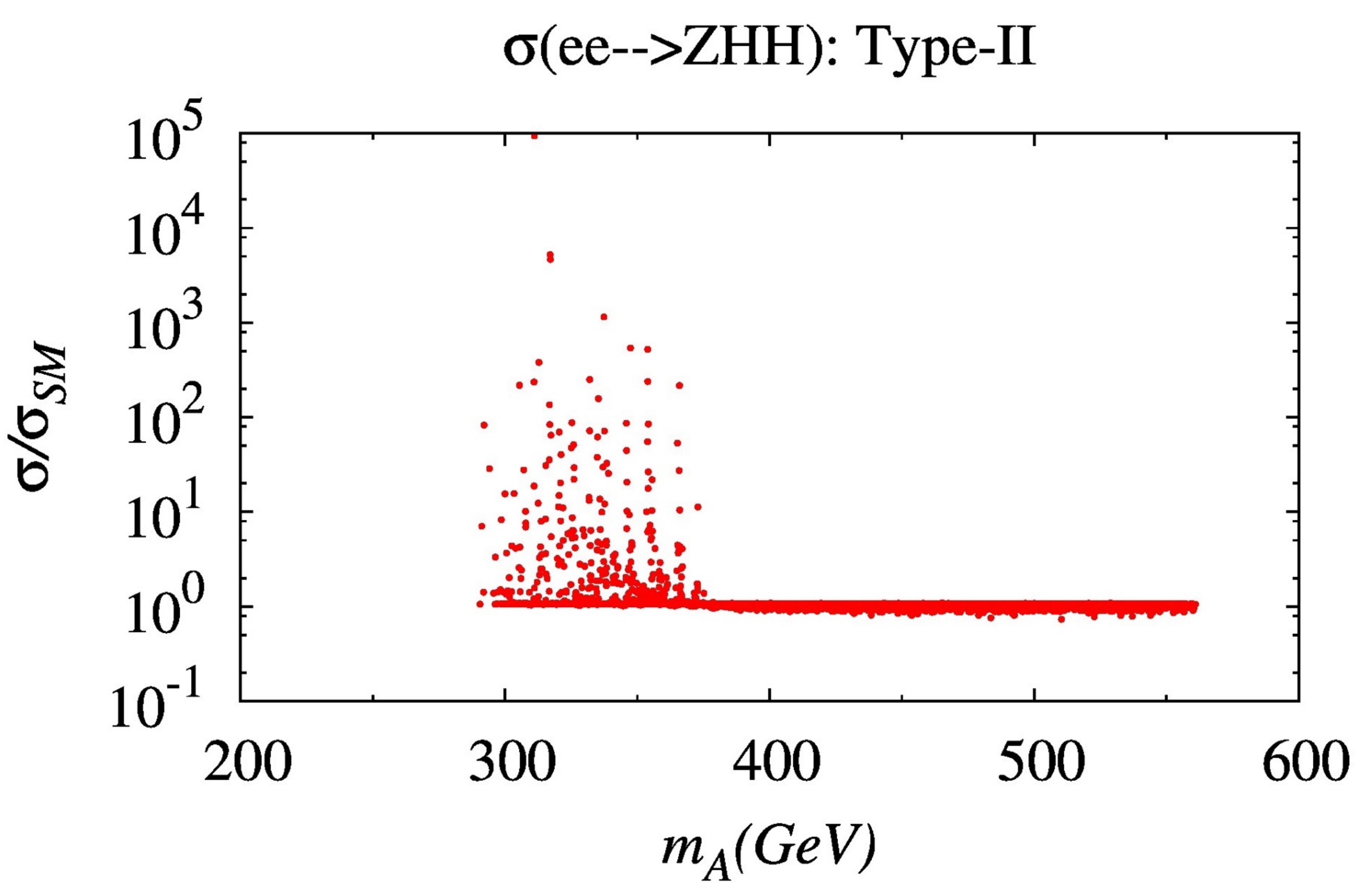}\\[0.3cm]
\includegraphics[width=8cm]{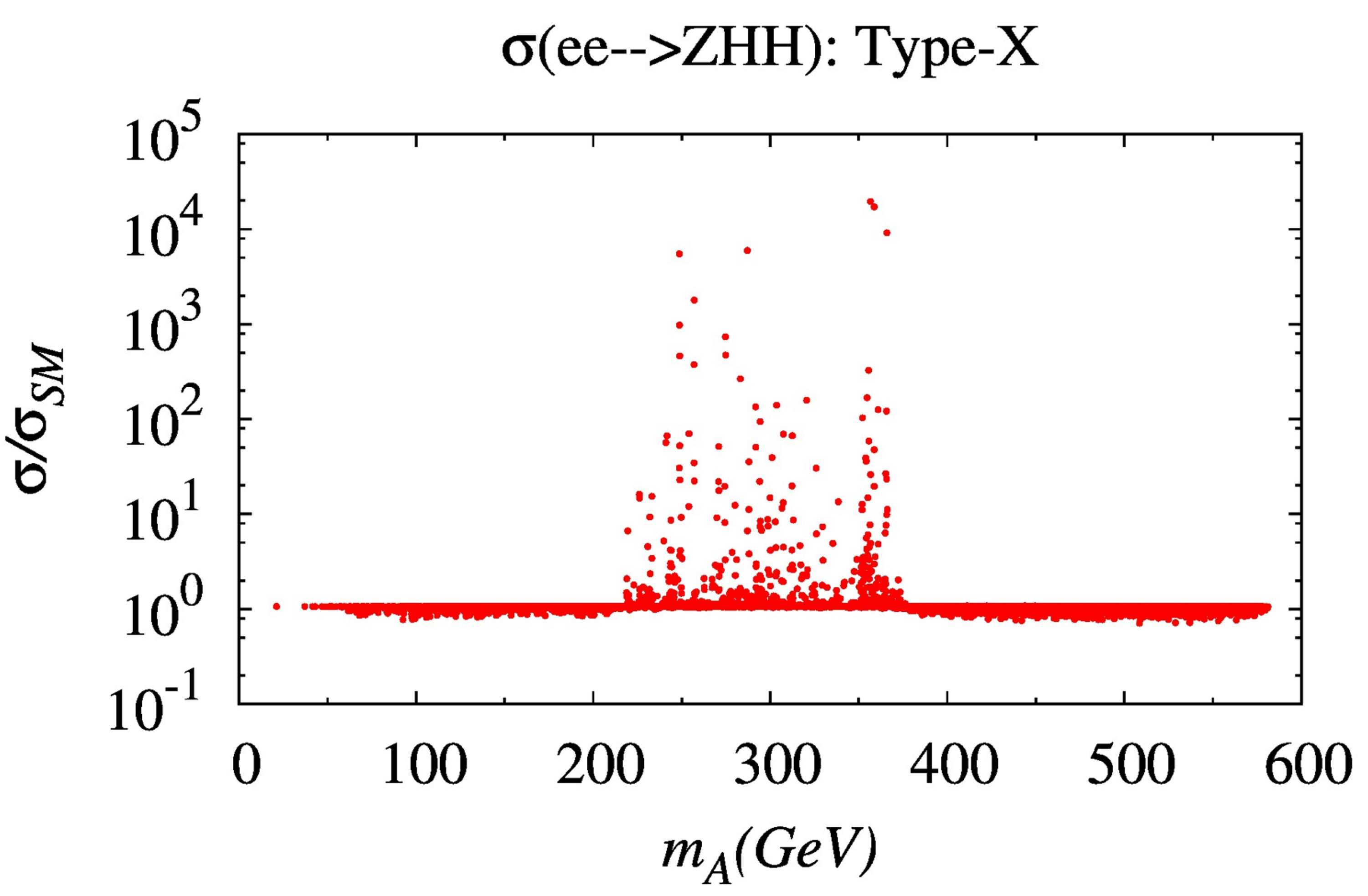}~~~~~
\includegraphics[width=8cm]{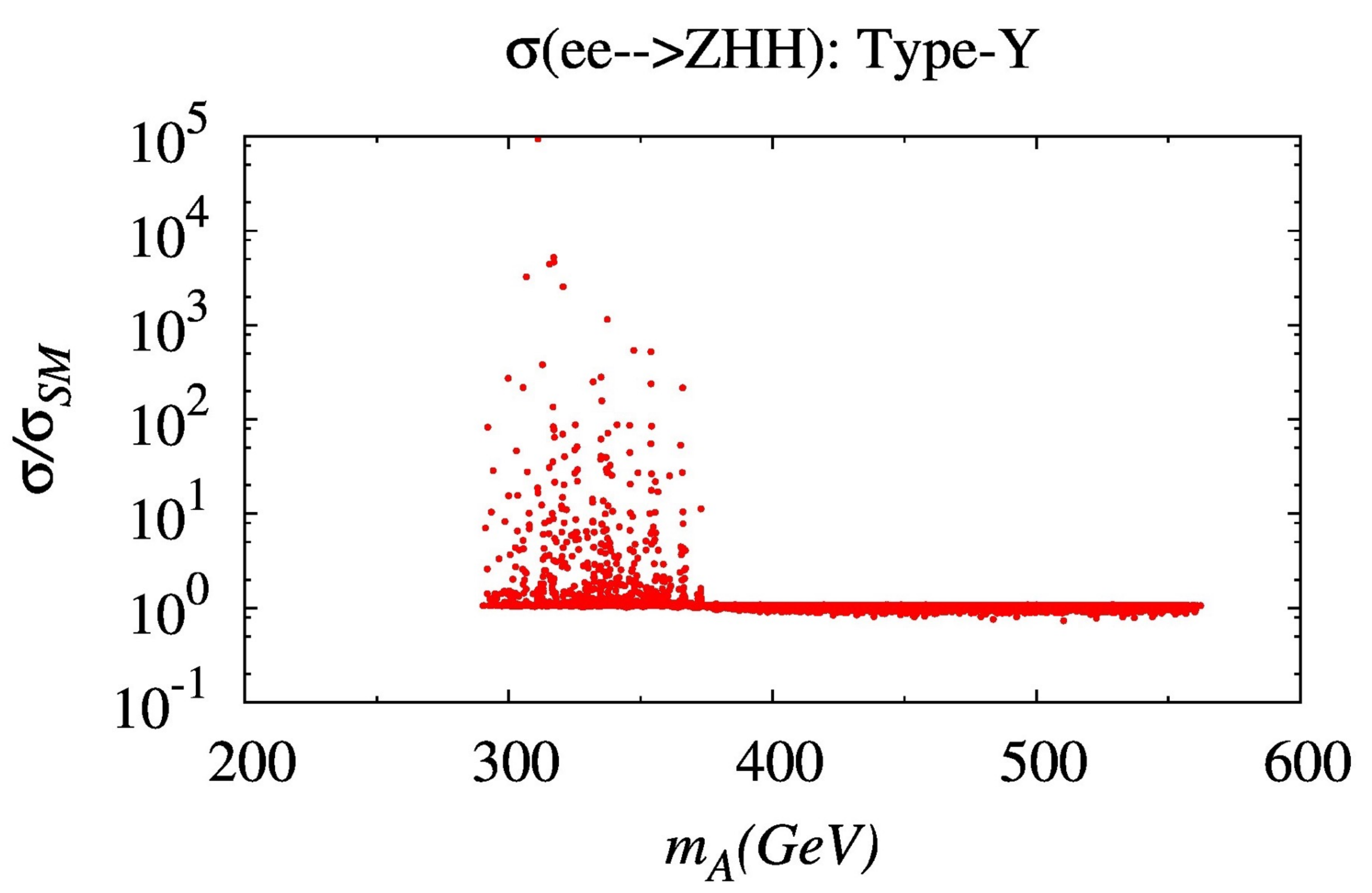}
\caption{\label{fig:eeZHHsig:mA}
The cross section of $e^+ e^- \to Z^0 H^0 H^0$
versus $m_A$ at $\sqrt{s}=500\gev$, normalized by the SM cross section, 
for the parameter points satisfying the LHC 
Higgs data.
 }
\end{figure}

Figure \ref{fig:eeZHHsig:mA} shows the expected 
total cross section of $e^+ e^- \to Z^0 H^0 H^0$
versus $m_A$ at $\sqrt{s}=500\gev$, normalized by the SM cross section.
We accept only the parameter points which satisfy all of the current constraints.
The analytic expression for $\sg(e^+ e^- \to Z^0 H^0 H^0)$
is referred to Ref.~\cite{Djouadi:1999gv}.
In most parameter space, the cross section in the 2HDM is
very similar to that in the SM.
In some parameter space, however,
 all of the four types allow highly enhanced cross section.
The rate of increase can be as large as factor of $10^4$.
As clearly shown in Types I and X, the enhancement occurs when $m_A \geq m_Z + m_H$.

\section{Conclusions}
\label{sec:conclusions}
We have investigated the meaning of the LHC Run 1 in the context of the hidden 
light Higgs scenario in the 2HDM with \textit{CP} invariance and the 
softly broken $Z_2$ symmetry. 
We found that the LHC Run 1 data combined with other current constraints do not exclude 
the possibility that the observed scalar particle 
is the heavier \textit{CP}-even Higgs boson $H^0$.
The lighter \textit{CP}-even Higgs boson $h^0$
is buried in the
region of $\lesssim 120$ GeV.
A remarkable consequence is that in order to make $m_H =125\gev$
the $Z_2$ symmetry breaking parameter $m_{12}^2$ 
cannot be large,  
which renders $m_{A, H^\pm}$ rather light at the sub-TeV scale.
We found the upper bounds on $m_{A, H^\pm}$ to be around $600\gev$.
Since the mass scale of other Higgs bosons are not far from the LHC reach, 
the hidden light Higgs scenario can be tested in the near future.

We also found that the LHC Run 1 data begin to constrain the Higgs potential 
of the 2HDM.
In particular, the values of $\lambda_2$ and
$\hat{g}_{HHH}$ are almost determined.
The cross section of $e^+e^- \to Z^0 H^0H^0$ is expected to be close to that of the SM, 
while in a limited region of $m_A$ around 300 GeV it could be highly enhanced.
The Higgs quartic couplings are less constrained.
Hopefully future lepton colliders could check our predictions.

\acknowledgments
SKK was supported by the National Research Foundation of Korea, NRF-2014R1A1A2057665.
JPL was supported by the second stage of the Brain Korea 21 Plus Project in 2015.
JS was supported by the National Research Foundation of Korea, NRF-2013R1A1A2061331. 
We thank Convergence Computing team of National Institute for Mathematical
Sciences for valuable comments in extracting data from experimental papers.


\end{document}